\documentclass[12pt]{report}
\usepackage{amsmath}
\usepackage{graphicx}
\usepackage{mathrsfs}

\begin{document}
\title{\Huge  \textbf{Decoherence dynamics of open qubit systems  }}
\author{\bf \\ \\ \\ \\ By\\ \\ \\ {\bf Qing-Jun Tong } \\ \\ \\ A thesis submitted in conformity with the requirements \\ for the degree of Master of philosophy\\
School of Nuclear Science and Technology, Lanzhou University}
\date{\today}
\maketitle
\newpage
\thispagestyle{empty}
 \ \ \ \

\newpage
\begin{center} {\Huge\textbf{Abstract}}\vskip0.7cm
\addcontentsline{toc}{chapter}{Abstract}
\end{center}
\parindent0.6cm

This thesis is contributed to the study of decoherence dynamics of
the dissipative qubit system. We mainly concentrate on the profound
impact of the formation of a bound state between the qubit and its
local environment on the decoherence behavior of the reduced qubit
system under the non-Markovian dynamics.

Firstly, we evaluate exactly the non-Markovian effect on the
decoherence dynamics of a single qubit interacting with a
dissipative vacuum reservoir. We find that the quantum coherence of
the qubit can be partially trapped in the steady state when the
non-Markovian memory effect of the reservoir is taken into account.
Our analysis shows that it is the formation of a bound state between
the qubit and its reservoir that results in this residual coherence
in the steady state under the non-Markovian dynamics. A physical
condition for the formation of the bound state is given explicitly.
Our results suggest a potential way to decoherence control by
modifying the system-reservoir interaction and the spectrum of the
reservoir to the non-Markovian regime in the scenario of reservoir
engineering.

Secondly, We study the entanglement dynamics of two qubits locally
interacting with their reservoirs and explore the entanglement
preservation under the non-Markovian dynamics. We show that the
existence of a bound state of the qubit and its reservoir and the
non-Markovian effect are two essential ingredients and their
interplay plays a crucial role to preserve the entanglement in the
steady state. When the non-Markovian effect is neglected, the
entanglement sudden death is reproduced. On the other hand, when the
non-Markovian is significantly strong but the bound state is absent,
the phenomenon of the entanglement sudden death and its revival is
recovered. Our formulation presents for the first time a unified
picture about the entanglement preservation and provides a clear
clue on how to preserve the entanglement in quantum information
processing.

Finally, in order to obtain a thorough understanding of the
entanglement dynamics, we study the entanglement distribution of a
two-qubit system, each of which is embedded into its local
reservoir, among all the bipartite subsystems including qubit-qubit,
qubit-reservoir, and reservoir-reservoir. Different to the result
that the entanglement of the qubits is transferred entirely to the
reservoirs under the Markovian dynamics, we find that the
entanglement can be stably distributed among all components under
the non-Markovian dynamics, and particularly it also satisfies an
identity firstly given by Y\"onac, Yu and Eberly [J. Phys. B
\textbf{40}, S45 (2007)] for a double J-C model without decoherence.
While the explicit distribution of the entanglement is dependent of
the detail of the model, even the approximation used, the identity
remains unchanged. Our unified treatment includes the previous
results in the literature as special cases. The result reveals the
profound nature of the entanglement and should have significant
implications for quantum information processing.

This thesis may give a clear clue of decoherence dynamics
under different approximations and how to preserve quantum coherence in
the steady state.

\newpage \pagenumbering{roman} \setcounter{page}{1} 
\textheight 210mm \textwidth 140mm \voffset -20mm
\footskip 15mm \markright{18} 
\pagestyle{plain} \newpage \pagenumbering{roman}
\setcounter{page}{1}

\tableofcontents \pagestyle{plain}

\newpage \pagenumbering{arabic}

\baselineskip 18pt


\chapter{Introduction}

The superposition rule of quantum state, one of the fundamental
principles of quantum mechanics, allows a quantum system to be in a
linear and coherent superposition of all possible quantum states
\cite{Breuer02}. It leads to the quantum coherence, which is the
essential difference of quantum world to the classical one. Quantum
coherence is of great importance not only in understanding the basic
rules of quantum mechanics but also in quantum information science
\cite{Nielsen00}.

Entanglement (also named as quantum correlation), as a non-local
quantum coherence, is one of the characteristic trait of quantum
mechanics \cite{Breuer02}. A state in multipartite system is
entangled when it cannot be written as the summation of the product
states of the subsystems. On the one hand, entanglement relates to a
lot of fundamental problems in quantum mechanics, such as, reality,
local realism, hidden variables, and quantum measurement theory
\cite{EPR1935,Bell1964,Zurek}. On the other hand, entanglement can
be used as a kind of information resource to realize some missions
of quantum information processing which are intractable for
classical one, such as quantum communication
\cite{Bennett1992,Bennett1993,Pan98}, quantum computation
\cite{Shor1994,Grover}, and quantum cryptography \cite{Ekert1993}.

Any realistic quantum system inevitably interacts with its
surrounding environment which leads to the loss of quantum
coherence, or decoherence of the quantum system. This ubiquitous
phenomenon deteriorates the superposition and the entanglement of
quantum state. In terms of information or energy, decoherence means
that the information or energy flows from quantum system to the
environment irreversibly. The decoherence is deemed as one of the
main obstacles to the realization of quantum information processing.
Recently, much attention has been paid on the study of the dynamics
of open quantum system, by which people want to get a thorough
understanding to the detrimental effect caused by decoherence on
quantum information processing and some clues on how to suppress
this unwanted effect. In this thesis, we will concentrate on a
detailed study of the decoherence dynamics of qubits influenced by
their vacuum reservoirs under different environments or
approximations and explore the potential dynamical suppression
mechanism to the decoherence.

The dynamics of open quantum system is rather complicated because of
the complex structure of the environment with which the system of
interest interacts. Actually, the exactly solvable models are very
few, only including the quantum Brownian motion and the system of a
two-level atom in a vacuum reservoir with Lorentzian spectrum
density \cite{Breuer02}. Many approximations are ordinarily
performed. A generally used approximation in the conventional
investigation to the dynamics of open quantum system is the
Born-Markovian approximation \cite{Gardiner}, which treats the
interaction between the quantum system of interest and its
environment perturbatively and neglects the memory effect of the
environment. This approximation is valid when the coupling between
quantum system and its environment is weak (Born approximation) and
the environmental correlation time is small compared to the typical
time scale of the quantum system (Markovian approximation). This
approximation yields equations of motion such as Redfield or master
equation, which is local in time and mathematically tractable, for
the quantum system of interest. Based on this approximation, it is
widely accepted that the quantum coherence of a single quantum
system flows irreversibly to the environment and the decoherence
dynamics can be simply depicted as an exponential decay. However,
things are changed dramatically when entanglement dynamics which
involves more subsystems, such as qubits, in the quantum system, are
studied. some works have showed that the entanglement between two
qubits ceases abruptly in a finite time scale
\cite{Zyczkowski2001,Yu04}. This remarkable phenomenon that the
entanglement of the qubits under the Markovian decoherence dynamics
can be terminated in a finite time despite the coherence of single
qubit lossing in an asymptotical manner is named as entanglement
sudden death (ESD). Further investigation shows that such ESD is
strongly related to the initial portion of double excitation
component \cite{Ikram07}. The larger the initial portion of the
component is, the shorter the death time is. On the other hand if
the environments are composed of the thermal or squeezed reservoirs,
it is found that the ESD would always happen for any initial
entangled state. Experimentally, the ESD has been observed using an
all optical setup and atomic ensemble system
\cite{Almeida07,Laurat07}.

The Born-Markovian approximation simplifies greatly the mathematics
to solve the dynamics of open quantum system, but it suffers more
and more challenges under the newly emerging experimental results
\cite{Fujita05,Dubin07,Koppens07,Mogilevtsev08,Xu09}. Especially,
when the environment has certain structures, such as atom in cavity
or photonic band gap (PBG) mediums
\cite{Scala08,Kaer10,Yablonovitch87,John1994,Lambropoulos00,Bellomo08},
the non-Markovian effect can not be neglected anymore. The
non-Markovian effect is a kind of dynamic feedback effect which
arises from the memory effect of environment.  In terms of
information or energy, the non-Markovian effect means that the
information or energy flows back from the environment to the quantum
system of interest. The study of open quantum dynamics beyond the
Markovian approximation is rather complicated, which needs the
solving of coupled integro-differential equations. The decoherence
dynamics of quantum system in this case exhibits a dramatic
deviation from the exponential decay behavior. Actually any kind of
environment should have memory effect. When this memory effect is
very weak, the Markovian approximation is applicable. On the other
hand, when the memory effect of the environment is extremely strong,
it would partially feed the lost coherence back to the quantum
system. In this case the Markovian approximation would be not
applicable. As far as the entanglement in two-qubit system is
concerned, the non-Markovian effect also has a great impact on it.
Modeling the environments as vacuum lossy cavities, Bellomo {\it et
al.} showed that the entanglement would revive again after a finite
period time of completely disappearance \cite{Bellomo07}. This is a
solvable model and the entanglement dynamics can be analytically
expressed. Via tuning relevant parameters, one can easily observe
that the non-Markovian effect postpones greatly the death of the
qubits entanglement. Entanglement dynamics of continues variable
system has also been well studied and the non-Markovian effect also
makes the entanglement dynamics oscillate
\cite{Goan07,An07,An092,An08,An09}, which can be understood as the
backaction effect of the environment on the quantum system. The ESD
and its revival due to the non-Markovian effect has been
experimentally observed \cite{Xu09}. All these experimental and
theoretically works show clearly that the coherence or entanglement
time of the quantum system can be much enhanced by the non-Markovian
effect.

However, in many cases such finite extension of the
coherence/entanglement time is not enough for quantum information
processing and thus it is desired to preserve a significant of the
quantum coherence/entanglement, even partially, in the long time
limit forever. Actually, some work has shown that it is realizable
for some special environment cases. It has been found that the
spontaneous emission of a two-level atom can be inhibited and its
quantum coherence can be preserved when the atom is placed in a PBG
material \cite{John1994,John1990,John99}. In the PBG material, the
photonic mode density is zero within the PBG and this would be
accompanied by the classical light localization and a photon-atom
bound state. The excited-state population in this case is partially
trapped, a phenomenon known as population trapping
\cite{Lambropoulos00}. This result has been verified experimentally
for quantum dot embedded in PBG material \cite{Lodahl04}. It has
been realized that trapping the single-qubit population is the key
step to protect entanglement in two-atom case \cite{Bellomo08}. When
two initially entangled qubits are immersed in two separate PBG
mediums, the entanglement of the two qubits can be preserved in the
steady state with a large fraction. However the mechanism of
entanglement preservation is still unclear. Also is this a general
phenomenon in open quantum system or only available in this specific
structured environment still is an open question.

To explore these questions, one should know the dynamics of quantum
system not only in the short-time scale, but also in the long-time
situation. In the short-time scale, when the non-Markovian effect is
very strong, the quantum coherence would surfer transient
oscillations manifesting the backaction effect of the memory
environment. It is just the counteraction role played by this
backaction effect to the dissipation effect of the environment on
the dynamics of the quantum system which results in the residual
coherence of the quantum system in the long-time limit. Therefore,
it is understandable that the non-Markovian effect is a prerequisite
for the coherence preservation in the long-time limit. Then a
natural question is: under the non-Markovian dynamics, what is the
condition for the quantum coherence to be preserved in the long-time
limit? This reminds us to examine the eigen solution of the whole
system, which actually determines the long-time behaviors of quantum
system. Firstly, let's consider the special case: the environment
contains only one mode, which corresponds to the J-C model. The
whole system possesses two real eigenvalues for each
excitation-number subspace. Consequently, the quantum coherence of
the two-level atom would experience loseless oscillations when the
degree of freedom of the single-mode environment is traced out.
However, when the environment possesses infinite modes, this would
not be the truth anymore. It was shown that the real eigenvalues are
not available anymore (except for the trivial ground state with
eigenvalue being zero) and the complex eigenvalues are present when
the environment has infinite modes \cite{John1990,Miyamoto}. This is
understandable based on the fact that the decay behavior of the
quantum system under decoherence is just the effect taken by the
imaginary part of the complex eigenvalues on the dynamics of reduced
quantum system. However, John \textit{et al} found that there is a
real eigenvalue available when the environment is a PBG medium
\cite{John1990}. Physically, the existence of a real eigenvalue
means the formation of atom-photon bound state. Due to the formation
of bound state, spontaneous emission would be suppressed and a large
proportion of quantum coherence may be preserved in the long-time
limit. And Ref. \cite{Bellomo08} reported that the entanglement can
be preserved with a large proportion in the long-time limit when two
atoms are embedded in the PBG mediums. We argue that the suppression
of the spontaneous emission and the entanglement preservation are
both contributed from the formation of the bound state. Is this
bound state available only for such PBG environment or for any
environment? Under what condition the coherence or entanglement
preservation is available for generic environments? These questions
motivate us to do the investigation in Chapters \ref{sqdq} and
\ref{meptq}.

There is always lots of entanglement of the quantum system lost
irrespective of the entanglement could be (partially) preserved or
not. Then a nature question is: where does the lost entangle go?
Modeling the whole systems as double J-C model, authors in Ref.
\cite{Yonac} have shown that the entanglement oscillates between the
atoms and the cavities in a lossless way. This is understandable
since there is no decoherence in the J-C model. Via introducing a
normalized collective state, authors in Ref. \cite{Lopez08} showed
that the initial entanglement between the qubits flows entirely to
their local environments under the Markovian dynamics. We argue that
things would be completely different under the non-Markovian
decoherence dynamics. This judgement is based on the following
observations. Firstly, it is possible to preserve some entanglement
in the quantum system under the non-Markovian dynamics, which means
that not all of the entanglement between the subsystems is
transferred to their environments. Secondly, the entanglement
preservation is due to the formation of the bound state between each
subsystem and its environment. Therefore, entanglement would exist
between each subsystem and its local environment. From these facts,
we can see that the question where does the entanglement go should
be reevaluated when the non-Markovian effect is taken into account.
This motivates us to do the investigation in Chapter \ref{edvtq}.

This thesis is organized as follows. In Chapter \ref{sqdq}, the
decoherence dynamics of single qubit is studied. We investigate the
exact decoherence dynamics of a dissipative qubit coupling to a
vacuum reservoir. We also study the static eigenvalue problem and
give the condition when atom-reservoir bound state is formed, via
which we reveal the mechanism of dynamical decoherence suppression
due to the bound state. In Chapter \ref{meptq}, entanglement
dynamics of two qubits under the influence of two independent vacuum
reservoirs is studied. We give a mechanism of entanglement
preservation. In Chapter \ref{edvtq}, we study the entanglement
distribution among all possible bipartite partitions of the same
system. Finally, a summary of this thesis and the outlook of future
works are given in Chapter \ref{smotl}.

\chapter{Decoherence dynamics of a dissipative qubit}\label{sqdq}

In this chapter we study the exact decoherence dynamics of a single
qubit (two-level atom) in a vacuum reservoir. We compare this result
with the one obtained under the Markovian approximation. We also
study the formation of bound state on the decoherence suppression.

To solve the dynamics for the general open quantum system is rather
tricky. Here we consider that the environment, with which the qubit
interacts, is in a vacuum state initially. Combining with numerical
calculations, we can obtain the exact decoherence dynamics of the
dissipative qubit.

\section{Introduction}

Any realistic quantum system inevitably interacts with its
surrounding environment, which leads to the loss of coherence, or
decoherence, of the quantum system \cite{Breuer02}. The decoherence
of quantum bit (qubit) is deemed as a main obstacle to the
realization of quantum computation and quantum information
processing \cite{Nielsen00}. Understanding and suppressing the
decoherence are therefore a major issue in quantum information
science. For a Markovian environment, it is well known that the
coherence of a qubit experiences an exponential decrease
\cite{Breuer02}. To beat this unwanted degradation, many controlling
strategies, passive or active, have been proposed
\cite{decohControl1,decohControl2,decohControl3,decohControl4,decohControl5}.

In recent years much attention has been paid to the non-Markovian
effect on the decoherence dynamics of open quantum system
\cite{Garraway97, Maniscalco2006, Pillo08, Koch08,
Breuer08,Rebentrost09,Breuer2009,Chruscinski10}. The significance of
the non-Markovian dynamics in the study of open quantum system is
twofold. i) It is of fundamental interest to extend the
well-developed methods and concepts of Markonian dynamics to
non-Markovian case \cite{Breuer02, Gardiner} for the open quantum
system in its own right. ii) There are many new physical situations
in which the Markovian assumption usually used is not fulfilled and
thus the non-Markovian dynamics has to be introduced. In particular,
many experimental results have evidenced the existence of the
non-Markovian effect
\cite{Fujita05,Dubin07,Koppens07,Mogilevtsev08}, which indicates
that one can now approach the non-Markovian regime via tuning the
relevant parameters of the system and the reservoir. The
non-Markovian effect means that the environment, when its state is
changed due to the interaction with the quantum system, in turn,
exerts its dynamical influence back on the system. Consequently one
can expect decoherence dynamics of the quantum system could exhibit
a dramatic deviation from the exponential decaying behavior. In
2005, DiVincenzo and Loss studied the decoherence dynamics of the
spin-boson model for the Ohmic heat bath in the weak-coupling limit.
They used the Born approximation and found that the coherence
dynamics has a power-law behavior at long-time scale \cite{Loss05},
which greatly prolongs the coherence time of the quantum system.
Such power-law behavior suggests that the non-Markovian effect may
play a constructive role in suppressing decoherence of the system.
Nevertheless, in many cases the finite extension of the coherence
time of the system is not sufficient for the quantum information
processing, a question arises whether the coherence of the system
can be preserved in the long-time limit, even partially.
Theoretically, the answer is positive if the environment has a
nontrivial structure. It has been shown that some residual coherence
can be preserved in the long-time steady state when the environment
is a periodic band gap material
\cite{Yablonovitch87,John1994,Lambropoulos00,Bellomo08} or leaky
cavity \cite{Scala08,Kaer10}. It is stressed that the residual
coherence is due to the confined structured environment. A natural
question is: Whether the coherence of the system can be dynamically
preserved or not by the non-Markovian effect if the environment has
no any special structure, e.g., a vacuum reservoir?

In this chapter, we study the exact decoherence dynamics of a qubit
interacting with a vacuum reservoir and examine the possibility of
decoherence suppression using the non-Markovian effect. The main aim
of this chapter is to analyze if and how the coherence present in the
initial state can be trapped with a noticeable fraction in the
steady state even when the environment is consisted of a vacuum
reservoir with trivial structure. We show that the non-Markovian
effect manifests its action on the qubit not only in the transient
dynamical process, but also in the asymptotical behavior. Our
analysis shows that the physical mechanism behind this dynamical
suppression to decoherence is the formation of a bound state between
the qubit and the reservoir. The no-decaying character of the bound
state leads to the inhibition of the decoherence and the residual
coherence trapped in the steady state. A similar vacuum induced
coherence trapping in the continuous variable system has been
reported in \cite{An08,An09}. Such coherence trapping phenomenon
provides an alternative way to suppress decoherence. This could be
realized by controlling and modifying the system-reservoir
interaction and the properties of the reservoir
\cite{Yablonovitch87} by the recently developed reservoir
engineering technique \cite{Wineland001,Wineland002,Diehl08}.

\section{The model and exact decoherence dynamics of the qubit}

We consider a qubit (two-level atom) which interacts with a vacuum
quantized radiation electromagnetic field. The Hamiltonian of the
total system reads
\cite{Breuer02}%
\begin{equation}
H=\omega _{0}\sigma _{+}\sigma _{-}+\sum_{k}\omega _{k}a_{k}^{\dag
}a_{k}+\sum_{k}(g_{k}\sigma _{+}a_{k}+h.c.),  \label{t2.1}
\end{equation}%
where $\omega _{0}$ is the transition frequency and $\sigma _{\pm }$
is the raising and lowering operators of the qubit and $a_{k}^{\dag
}$ and $a_{k}$,
respectively, are the creation and annihilation operators of the \textbf{k}%
-th mode with frequency $\omega _{k}$ of the radiation field. The
coupling
strength between the qubit and the radiation field is given by%
\begin{equation}
g_{k}=-i\sqrt{\frac{\omega _{k}}{2\varepsilon _{0}V}}\mathbf{\hat{e}}%
_{k}\cdot \mathbf{d,}  \label{t19}
\end{equation}%
where $\mathbf{\hat{e}}_{k}$ and $\mathbf{d}$ are unit polarization
vector of the radiation field and the dipole moment of the qubit
respectively. Thoughout this paper we assume $\hbar =1$.

To obtain the exact dynamics of the qubit, we first consider the
following two simple cases. For simplicity, we assume there is no
correlation between the qubit and its reservoir at the initial time
$t=0$. If the initial state of the system is $\left\vert \Psi
(0)\right\rangle =\left\vert -,\{0_{k}\}\right\rangle $, where
$\left\vert -\right\rangle $ denotes the ground state of the qubit
and $\left\vert \{0_{k}\}\right\rangle $ represents the vacuum state
of the reservoir, the whole system will not evolve with time under
the Hamiltonian (\ref{t2.1}). Whereas to the initial state
$\left\vert \Psi (0)\right\rangle =\left\vert
+,\{0_{k}\}\right\rangle $, in which $\left\vert +\right\rangle $
denotes the exited state of the
qubit, the time evolution of the total system has the following form%
\begin{equation}
\left\vert \Psi (t)\right\rangle =b_{0}(t)\left\vert
+,\{0_{k}\}\right\rangle +\sum_{k}b_{k}(t)\left\vert
-,\{1_{k}\}\right\rangle ,
\end{equation}%
where $\left\vert \{1_{k}\}\right\rangle $ represents the field
state
containing one photon in the \textbf{k}-th mode. Applying the Schr\"{o}%
dinger equation, we get the time evolution of the probability amplitudes%
\begin{eqnarray}
i\dot{b}_{0}(t) &=&b_{0}(t)\omega _{0}+\sum_{k}g_{k}b_{k}(t),  \label{t2} \\
i\dot{b}_{k}(t) &=&b_{k}(t)\omega _{k}+g_{k}^{\ast }b_{0}(t),
\label{t3}
\end{eqnarray}%
where the superscript dot represents the differential with respect
to time.
Solving Eq. (\ref{t3}) formally and substituting the solution into Eq. (\ref%
{t2}), we can obtain%
\begin{equation}
\dot{b}_{0}(t)+i\omega _{0}b_{0}(t)=-\int_{0}^{t}f(t-\tau
)b_{0}(\tau )d\tau , \label{t4}
\end{equation}%
where the kernel function is $f(x)=\sum_{k=0}^{\infty }\left\vert
g_{k}\right\vert ^{2}\exp (-i\omega _{k}x)$. Obviously the memory
effect has been registered in the kernel function. In the continuous
limit of the
environment frequency, \ the kernel function has the form%
\begin{equation}
f(x)=\int_{0}^{\infty }J(\omega )e^{-i\omega x}d\omega , \label{crf}
\end{equation}%
where $J(\omega )=\eta \omega ^{3}e^{\frac{-\omega }{\omega _{c}}}$
is the spectral density \cite{Leggett87}, which characterizes the
coupling strength
of the reservoir to the qubit with respect to the reservoir frequency and $%
\eta =\frac{\int \left\vert \mathbf{\hat{e}}_{k}\cdot
\mathbf{d}\right\vert ^{2}d\Omega }{(2\pi c)^{3}2\varepsilon _{0}}$.
To eliminate infinity in
frequency integration, we have introduced the cutoff frequency $\omega _{c}$%
. On physical grounds, the introducing of the cutoff frequency means
that not all of the infinite modes of the reservoir contribute to
the interaction with the qubit, and one always expects the spectral
density going to zero for the modes with frequencies higher than
certain characteristic frequency. It is just this characteristic
frequency which determines the specific behavior and the properties
of the reservoir. One can see that in our model, the spectral
density has a super-Ohmic form \cite{Leggett87}.

From the time evolution of the above two situations, one can get the
time evolution of any given initial state of the system readily. For
an initially mixed state, which is described by the following
density operator
\begin{eqnarray}
\rho _{\text{tot}}(0) &=&(\rho _{11}\left\vert +\right\rangle
\left\langle +\right\vert +\rho _{12}\left\vert +\right\rangle
\left\langle -\right\vert
+\rho _{21}\left\vert -\right\rangle \left\langle +\right\vert \nonumber\\
&&+\rho _{22}\left\vert -\right\rangle \left\langle -\right\vert
)\otimes \left\vert \{0\}_{k}\right\rangle \left\langle
\{0\}_{k}\right\vert .
\end{eqnarray}%
The time evolution of the total system can be calculated explicitly.
In fact, what we care about is the reduced density matrix of the
qubit, which
is obtained by tracing over the reservoir variables%
\begin{equation}
\rho (t)=\left(
\begin{array}{cc}
\rho _{11}\left\vert b_{0}(t)\right\vert ^{2} & \rho _{12}b_{0}(t) \\
\rho _{21}b_{0}^{\ast }(t) & 1-\rho _{11}\left\vert b_{0}(t)\right\vert ^{2}%
\end{array}%
\right) .  \label{t5}
\end{equation}%
Differentiating Eq. (\ref{t5}) with respect to time, we may obtain
the
equation of motion of the qubit%
\begin{eqnarray}
\dot{\rho}(t) &=&-i\frac{\Omega (t)}{2}[\sigma _{+}\sigma _{-},\rho (t)]+%
\frac{\gamma (t)}{2}[2\sigma _{-}\rho (t)\sigma _{+}  \notag \\
&&-\sigma _{+}\sigma _{-}\rho (t)-\rho (t)\sigma _{+}\sigma _{-}],
\label{t6}
\end{eqnarray}%
where $\Omega (t)=-2\text{Im}[\frac{\dot{b}_{0}(t)}{b_{0}(t)}]$ and
$\gamma (t)=-2\text{Re}[\frac{\dot{b}_{0}(t)}{b_{0}(t)}]$. $\Omega
(t)$ plays the role of time-dependent shifted frequency and $\gamma
(t)$ that of time-dependent decay rate \cite{Breuer02}. It is worth
mentioning that during the derivation of master equation (\ref{t6})
we have not resorted to the Born-Markovian approximation. Therefore
Eq. (\ref{t6}) is the exact master equation of the qubit system.

It is interesting to notice that one can reproduce the conventional
Markovian one from our exact non-Markovian master equation under
certain approximations. By redefining the probability amplitude as
$b_{0}(\tau
)=b_{0}^{\prime }(\tau )e^{-i\omega _{0}\tau }$, one can recast Eq.~(\ref{t4}%
) into
\begin{equation}
\dot{b}_{0}^{\prime }(t)+\int_{0}^{\infty }d\omega J(\omega
)\int_{0}^{t}d\tau e^{i(\omega _{0}-\omega )(t-\tau )}b_{0}^{\prime
}\left( \tau \right) =0,  \label{t13}
\end{equation}%
where $J(\omega )$ is defined the same as above. Then, we take the
Markovian approximation,
\begin{equation}
b_{0}^{\prime }\left( \tau \right) \cong b_{0}^{\prime }(t),
\label{t14}
\end{equation}%
namely, approximately taking the dynamical variable to the one that
depend only on the present time so that any memory regarding the
earlier time is ignored. The Markovian approximation is mainly based
on the physical assumption that the correlation time of the
reservoir is very small compared with the typical time scale of
system evolution. Also under this assumption we can extend the upper
limit of the $\tau $ integration in Eqs.~(\ref{t13}) to infinity and
use the equality
\begin{equation}
\lim_{t\rightarrow \infty }\int_{0}^{t}d\tau e^{\pm i(\omega
_{0}-\omega )(t-\tau )}=\pi \delta (\omega -\omega _{0})\mp
iP\Big(\frac{1}{\omega -\omega _{0}}\Big),
\end{equation}%
where $P$ and the delta-function denote the Cauchy principal value
and the singularity, respectively. The integro-differential equation
in (\ref{t13}) is thus reduced to a linear ordinary differential
equation. The solutions of $b_{0}^{\prime }$ as well as $b_{0}$ can
then be easily obtained as
\begin{equation}
b_{0}(t)=e^{-i(\omega _{0}-\delta \omega )t-\pi J(\omega _{0})t},
\end{equation}%
where $\delta \omega =P\int_{0}^{\infty }\frac{J(\omega )d\omega
}{\omega -\omega _{0}}$. Thus one can verify that,
\begin{equation}
\gamma (t)\equiv \gamma _{0}=2\pi J(\omega _{0}),~~\Omega (t)\equiv
\Omega _{0}=2(\omega _{0}-\delta \omega ),  \label{t18}
\end{equation}%
which are exactly the coefficients in the Markovian master equation
of the two-level atom system \cite{Breuer02}.

\section{Purity and decoherence factor}\label{quant}

To quantify the decoherence dynamics of the qubit, we introduce the
following two quantities. The first one is the purity, which is
defined as \cite{Nielsen00}
\begin{equation}
p(t)=\text{Tr}\rho ^{2}(t).  \label{prtsp}
\end{equation}%
Clearly $p=1$ for pure state and $p<1$ for mixed state. The second
quantity describing the decoherence is the decoherence factor $c(t)$
of the qubit, which is determined by the off-diagonal elements of
the reduced density matrix
\begin{equation}
\left\vert \rho _{12}(t)\right\vert =c(t)\left\vert \rho
_{12}(0)\right\vert .  \label{chr12}
\end{equation}
The decoherence factor maintains unity when the reservoir is absent
and vanishes for the case of completely decoherence.

For definiteness, we consider the following initial pure state of
the qubit
\begin{equation}
\left\vert \psi (0)\right\rangle =\alpha \left\vert +\right\rangle
+\beta \left\vert -\right\rangle ,  \label{initsp}
\end{equation}%
in which $\alpha $ and $\beta $ satisfy the normalization condition.
Using Eq. (\ref{t5}), the exact time evolution of the qubit is
easily
obtained%
\begin{equation}
\rho (t)=\left(
\begin{array}{cc}
\left\vert \alpha \right\vert ^{2}\left\vert b_{0}(t)\right\vert
^{2} &
\alpha \beta ^{\ast }b_{0}(t) \\
\alpha ^{\ast }\beta b_{0}^{\ast }(t) & 1-\left\vert \alpha
\right\vert
^{2}\left\vert b_{0}(t)\right\vert ^{2}%
\end{array}
\right) .  \label{rhocom}
\end{equation}
With Eq. (\ref{rhocom}), the purity and decoherence factor can be
expressed explicitly
\begin{equation}
p(t)=2\left\vert \alpha \right\vert ^{4}\left\vert
b_{0}(t)\right\vert ^{2}[\left\vert b_{0}(t)\right\vert ^{2}-1]+1,
\label{tt11}
\end{equation}
and
\begin{equation}
c(t)=\left\vert b_{0}(t)\right\vert .  \label{tt12}
\end{equation}

It is easy to verify, under the Born-Markovian approximation, the
purity and
decoherence factor have the following forms%
\begin{equation}
p(t)=2\left\vert \alpha \right\vert ^{4}e^{-\gamma _{0}t}(e^{-\gamma
_{0}t}-1)+1,  \label{tt16}
\end{equation}
and
\begin{equation}
c(t)=e^{-\frac{\gamma _{0}}{2}t},  \label{tt17}
\end{equation}
where the time-independent decay rate $\gamma_0$ is given in Eq.
(\ref{t18}). Obviously, the system asymptotically loses its quantum
coherence ($c(\infty)=0$) and approaches a pure steady state
($p(\infty)=1$) irrespective of the form of the initial state under
the Markovian approximation. One can also find from Eqs.
(\ref{tt11}-\ref{tt17}) that the probability amplitude of excited
state plays key role in the decoherence dynamics.

\section{Numerical results and analysis}\label{numa}
\begin{figure*}[tbp]
\centering
\includegraphics[width = \columnwidth]{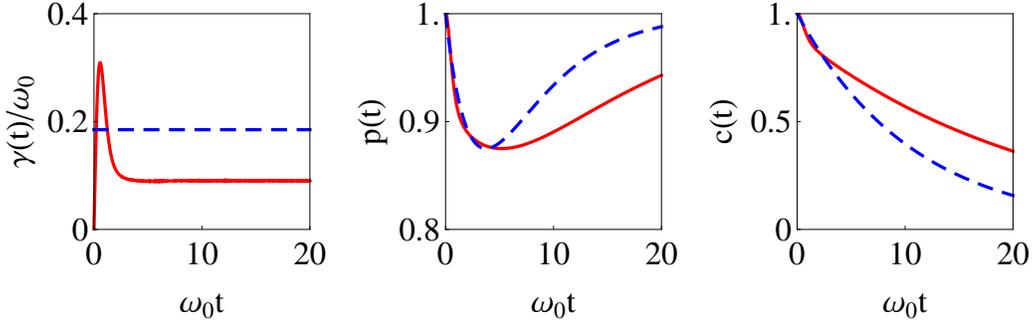}
\caption{ Time evolution of $\gamma (t)$, $p(t)$ and $c(t)$ in
non-Markovian situation (solid line) and the corresponding Markovian
situation (dashed line), when $\eta$ and $\omega_c$ are small. The
parameters used here are $\alpha =1/\sqrt{2}$, $\eta =0.08$ and
$\omega _{c}/\omega _{0}=1.0.$}\label{weakf}
\end{figure*}

In this section, by numerically solving Eq. (\ref{t4}), we study the
influence of memory effect of reservoir on the exact dynamics of the
qubit. Noticing the fact that the memory effect registered in the
kernel function is essentially determined by the spectrum density
$J(\omega )$, one can expect that $J(\omega)$ plays an major role in
the exact dynamics of the qubit. In the following, we show how the
decoherence of the qubit can be fully suppressed under the
non-Markovian dynamics in terms of the relevant parameters of
$J(\omega)$ \cite{tong1}.

\subsection{The influence of coupling constant}

In the following, we numerically analyze the exact decoherence
dynamics of the qubit with respect to decay rate $\gamma (t)$,
purity $p(t)$ and decoherence factor $c(t)$ in terms of the coupling
constant $\eta $ \cite{tong1}.

In Fig. \ref{weakf} we plot the time evolution of decay rate
$\gamma(t)$, purity $p(t)$, decoherence factor $c(t)$ and their
Markovian correspondences in the weak coupling and low cutoff
frequency case. We can see that $\gamma(t)$ shows distinct
difference from its Markovian counterpart over a very short time
interval. With time, $\gamma(t)$ tends to a definite positive value.
The small ``jolt" of $\gamma(t)$ in the short time interval just
evidences the backaction of the memory effect of the reservoir
exerted on the qubit \cite{Hu92}. It manifests that the reservoir
does not exert decoherence on the qubit abruptly, just as the result
based on Markovian approximation, but dynamically influences the
qubit and gradually establishes a stable decay rate to the qubit.
Furthermore, it is also shown that the decay rate is positive in the
full range of evolution, which results in any initial qubit state
evolving to the ground state $\left\vert \psi (\infty)\right\rangle
=\left\vert -\right\rangle $ irreversibly. Consequently the
decoherence factor monotonously decreases to zero with time and the
purity approaches unity in the long-time limit, which is consistent
with the result under Markovian approximation. The result indicates
that although the reservoir has backaction effect on the qubit, it
is quite small. And the dissipation effect of the reservoir
dominates the dynamics of the qubit. Thus no qualitative difference
can be expected between the exact result and the Markovian one with
the backaction effect ignored. Therefore the widely used Markovian
approximation is applicable in this case. Nevertheless, at the short
and immediate time scales the overall behavior is still quite
different from that of the Markovian dynamics. The decoherence
factor shown in the righ-hand panel of Fig. \ref{weakf} shows
non-exponential decay, which is in agreement with the result
obtained previously in the spin-boson model in the weak-coupling
limit \cite{Loss05}. However, the situation is dramatically changed
if the coupling is strengthened as discussed below.
\begin{figure*}[tbp]
\centering
\includegraphics[width = \columnwidth]{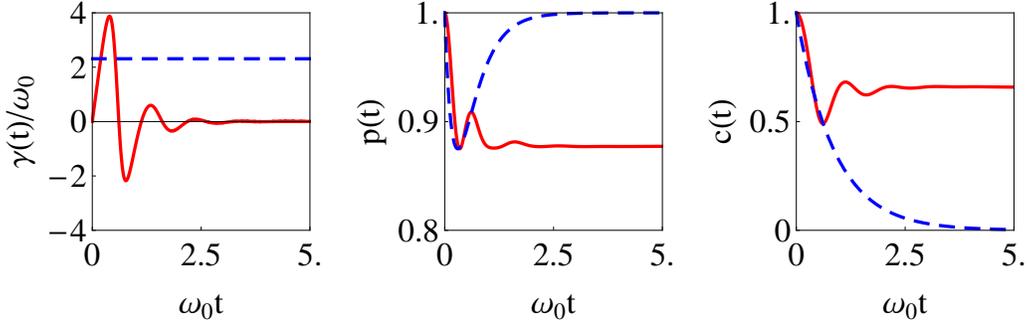}
\caption{Time evolution of $\gamma (t)$, $p(t)$ and $c(t)$ in
non-Markovian situation (solid line) and the corresponding Markovian
situation (dashed line), when $\eta$ is large. The parameters used
here are $\alpha =1/\sqrt{2}$, $\eta =1.0$ and $\omega _{c}/\omega
_{0}=1.0.$}\label{strongf}
\end{figure*}

With the same cutoff frequency as in Fig. \ref{weakf} but a larger
coupling constant, we plot in Fig. \ref{strongf} the decay rate,
purity and decoherence factor in the strong coupling case. In this
case the non-negligible backaction of the reservoir has a great
impact on the dynamics of the qubit. Firstly, we can see that the
decay rate not only exhibits oscillations, but also takes negative
values in the short time scale. Physically, the negative decay rate
is a sign of strong backaction induced by the non-Markovian memory
effect of the reservoir. And the oscillations of the decay rate
between negative and positive values reflect the exchange of
excitation back and forth between qubit and the reservoir
\cite{Pillo08}. Consequently both the decoherence factor and the
purity exhibit oscillations in a short-time scale, which shows
dramatic deviation to the Markovian result. Therefore, entirely
different to the weak coupling case in Fig. \ref{weakf}, the
reservoir in the strong coupling case here has strong backaction
effect on the qubit. Secondly, we also notice that the decay rate
approaches zero in the long-time limit. The vanishing decay rate
means, after several rounds of oscillation, the qubit ceases
decaying asymptotically. The non-Markovian purity maintains a steady
value asymptotically, which is less then unity. This indicates that
the steady state of the qubit is not the ground state anymore, but a
mixed state. The decoherence factor also tends to a non-zero value,
which implies that the coherence of the qubit is preserved with a
noticeable fraction in the long-time steady state. These phenomena,
which are qualitatively different to the Markovian situation,
manifest that the memory effect has a considerable contribution not
only to the short-time, but also to the long-time behavior of the
decoherence dynamics. The presence of the residual coherence in the
steady state also suggests a potential active control way to protect
quantum coherence of the qubit from decoherence via the
non-Markovian effect.
\begin{figure*}[tbp]
\centering
\includegraphics[width = \columnwidth]{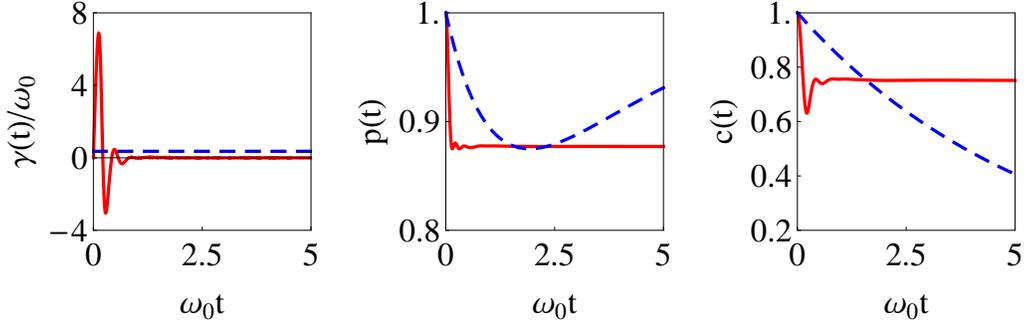}
\caption{Time evolution of $\gamma (t)$, $p(t)$ and $c(t)$ in
non-Markovian situation (solid line) and the corresponding Markovian
situation (dashed line), when $\omega _{c}$
is large. The parameters used here are $\alpha =1/\sqrt{2}$%
, $\eta =0.08$ and $\omega _{c}/\omega _{0}=3.0.$}\label{sc}
\end{figure*}

\subsection{The influence of cutoff frequency}

The cutoff frequency $\omega _{c}$, on the one hand, is introduced
to eliminate the infinity in the frequency integration. On the other
hand it also determines the frequency range in which the power form
is valid \cite{Weiss}. In the following, we elucidate the influence
of cutoff frequency on the exact decoherence dynamics \cite{tong1}.

Fixing $\eta$ as the value in Fig. \ref{weakf} and increasing the
cutoff frequency, we plot in Fig. \ref{sc} the dynamics of the qubit
in a high cutoff frequency case. It shows that a similar decoherence
behavior as the strong coupling case in Fig. \ref{strongf} can be
obtained. After several rounds of oscillation, the decay rate tends
to zero in the long-time limit. The negative decay rate makes the
lost coherence partially recovered. The vanishing decay rate in the
long-time limit results in the decoherence frozen before the qubit
gets to its ground state. Thus there is some residual coherence
trapped in the steady state. Similar to the strong coupling case, it
is essentially the interplay between the backaction and the
dissipation on the dynamics of qubit which results in the inhibition
of decoherence. We argue that in this high cutoff frequency regime,
the widely used Markovian approximation is not applicable because of
the strong backaction effect of the reservoir.

\subsection{The physical mechanism of the decoherence inhibition:
the formation of atom-photon bound state}

From the analysis above we can see clearly that the decoherence can
be inhibited in the non-Markovian dynamics. A natural question is:
What is physical mechanism to cause such dynamical decoherence
inhibition? To answer this question, let us find the eigen solution
of Eq. (\ref{t2.1}) in the sector of one-excitation in which we are
interested \cite{tong1}. The eigenequation reads $H\left\vert
\varphi _{E}\right\rangle =E\left\vert \varphi _{E}\right\rangle $,
where $\left\vert \varphi _{E}\right\rangle =c_{0}\left\vert
+,\{0_{k}\}\right\rangle +\sum_{k=0}^{\infty }c_{k}\left\vert
-,1_{k}\right\rangle $. After some algebraic calculation, we can
obtain a transcendental equation of $E$
\begin{equation}
y(E)\equiv\omega _{0}- \int_{0}^{\infty }\frac{J(\omega) }{\omega
-E}d\omega =E. \label{eigensp}
\end{equation}
From the fact that $y(E)$ decreases monotonically with the increase
of $E$ when $E<0$ we can say that if the condition $y(0)<0$, i.e.
\begin{equation} \omega _{0}-2\eta
\frac{\omega^3 _{c}}{\omega_0^2}<0\label{cdtsu}\end{equation} is
satisfied,
$y(E)$ always has one and only one intersection in the regime $%
E<0$ with the function on the right-hand side of Eq.
(\ref{eigensp}). Then the system will have an eigenstate with real
(negative) eigenvalue, which is a bound state \cite{Miyamoto,tong1},
in the Hilbert space of the qubit plus its reservoir. While in the
regime of $E>0$, one can see that $y(E)$ is divergent, which means
that no real root $E$ can make Eq. (\ref{eigensp}) well-defined.
Consequently Eq. (\ref{eigensp}) does not have positive real root to
support the existence of a further bound state. It is noted that Eq.
(\ref{eigensp}) may possess complex root. Physically this means that
the corresponding eigenstate experiences decay contributed from the
imaginary part of the eigenvalue during the time evolution, which
causes the excited-state population approaching zero asymptotically
and the decoherence of the reduced qubit system.

The formation of bound state is just the physical mechanism
responsible for the inhibition of decoherence. This is because a
bound state is actually a stationary state with a vanishing decay
rate during the time evolution. Thus the population probability of
the atomic excited state in bound state is constant in time, which
is named as ``population trapping"
\cite{Yablonovitch87,Lambropoulos00}. This claim is fully verified
by our numerical results. The parameters in Fig. \ref{weakf} do not
satisfy the condition (\ref{cdtsu}) to support the existence of a
bound state, then the dynamics experiences a severe decoherence.
While with the increase of either $\eta$ (in Fig. \ref{strongf}) or
$\omega_c$ (in Fig. \ref{sc}), the bound state is formed. Then the
system and its environment is so correlated that it causes the decay
rate of the system in the non-Markovian dynamics exhibiting: 1)
transient negative value due to the backaction of the environment;
2) vanishing asymptotic value. Such interesting phenomenon, i.e. the
vanishing asymptotical decay rate in the large cutoff frequency
regime for super-Ohmic spectrum density, was also revealed in Ref.
\cite{Paz09}. This effect of course is missing in the conventional
Born-Markovian decoherence theory, where the reservoir is
memoryless.

\begin{figure}[tbp]
\begin{center}
\includegraphics[scale=1]{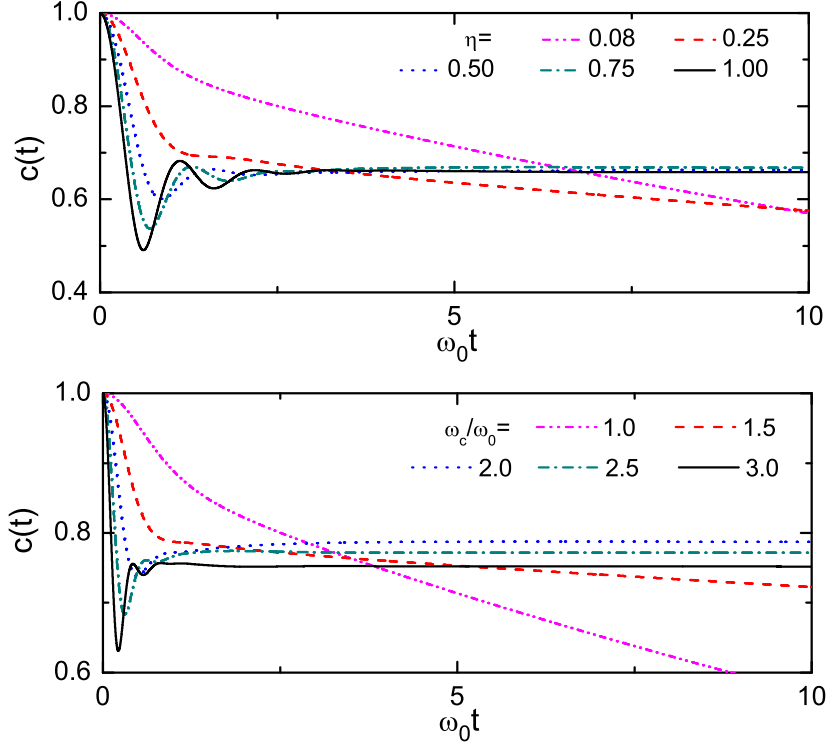}
\caption{Time evolution of $c(t)$ in the non-Markovian dynamics with
different $\eta$ when $\omega_c/\omega_0=1.0 $ (upper panel) and
with different $\omega_c$ when $\eta=0.08$ (lower
panel).}\label{comp}\end{center}
\end{figure}

In order to understand the exact decoherence dynamics more
completely, we plot in Fig. \ref{comp} the crossover from coherence
destroying to coherence trapping via increasing either the coupling
constant or the cutoff frequency. Coherence trapping can be achieved
as long as the bound state is formed. Therefore, one can preserve
coherence via tuning the relevant parameters of system and the
reservoir, e.g. the qubit-reservoir coupling constant and the
property of the reservoir so that the condition (\ref{cdtsu}) is
satisfied.

\section{Summary}\label{dsumma}
In summary, we have investigated the exact decoherence dynamics of a
qubit in a dissipative vacuum reservoir. We have found that even in
a vacuum environment without any nontrivial structure, we can still
get the decoherence suppression of the qubit owing to the dynamical
mechanism of the non-Markovian effect. From our analytic and
numerical results, we find that the non-Markovian reservoir has dual
effects on the qubit: dissipation and backaction. The dissipation
effect exhausts the coherence of the qubit, whereas the backaction
one revives it. In the strong coupling and/or high cutoff frequency
regimes, a bound state between the qubit and its reservoir is
formed. It induces a strong backaction effect in the dynamics
because the reservoir is strongly correlated with the qubit in the
bound state. Furthermore, because of the non-decay character of the
bound state the decay rate in this situation approach zero
asymptotically. The vanishing of the decay rate causes the
decoherence to cease before the qubit decays to its ground state.
Thus the qubit in the non-Markovian dynamics would evolve to a
non-ground steady state and there is some residual coherence
preserved in the long-time limit. Our results make it clear how the
non-Markovian effect shows its effects on the decoherence dynamics
in different parameter regimes.

The presence of such coherence trapping phenomenon actually gives us
an active way to suppress decoherence via non-Markovian effect. This
could be achieved by modifying the properties of the reservoir to
approach the non-Markovian regime via the potential usage of the
reservoir engineering technique
\cite{Wineland001,Wineland002,Diehl08,Garraway}. Many experimental
platforms, e.g. mesoscopic ion trap \cite{Wineland001,Wineland002},
cold atom BEC \cite{Diehl08}, and the photonic crystal material
\cite{Yablonovitch87} have exhibited the controllability of
decoherence behavior of relevant quantum system via well designing
the size (i.e. modifying the spectrum) of the reservoir and/or the
coupling strength between the system and the reservoir. It is also
worth mentioning that a proposal aimed at simulating the spin-Boson
model, which is relevant to the one considered in this paper, has
been reported in the trapped ion system \cite{Porras}. On the other
side many practical systems can now be engineered to show the novel
non-Markovian effect \cite{Dubin07,Koppens07,Mogilevtsev08,Xu09}.
All these achievements show that the recent advances have paved the
way to experimentally simulate the paradigmatic models of open
quantum system, which is one part of the new-emergent field, quantum
simulators \cite{Nori}. Our work sheds new light on the way to
indirectly control and manipulate the dynamics of quantum system in
this experimental platforms.

A final remark is that our results can be generalized to the system
consisted of two qubits, each of which interacts with a local
reservoir. Because of the coherence trapping we expect that the
non-Markovian effect plays constructive role in the entanglement
preservation \cite{Bellomo08,An09,Tong09}.

\chapter{Mechanism of entanglement preservation}\label{meptq}

In this chapter, we study the mechanism of entanglement
preservation. We found entanglement can be preserved in the long
time limit as long as bound states are formed in the local systems.
We also find the non-Markovian effect has a profound effect on the
entanglement preservation.

\section{Introduction}

Entanglement is not only of fundamental interest to quantum
mechanics, but also of great importance to quantum information
processing \cite{Nielsen00}. However, due to the inevitable
interaction of qubits with their environments, entanglement always
experiences degradation. A phenomenon that the entanglement between
two qubits may completely disappear at a finite time, known as
``entanglement sudden death" (ESD), has been predicted theoretically
\cite{Zyczkowski2001,Yu04} and subsequently been verified
experimentally \cite{Almeida07,Laurat07}, which indicates the
specific behavior of the entanglement different from the coherence.
From the point of view of applications, the ESD is apparently
disadvantageous to the quantum information processing.

Recently, Bellomo {\it et al.} \cite{Bellomo07} found that the
entanglement can revive after some time interval of the ESD and thus
extends significantly the entangled time of the qubits. This
remarkable phenomenon, which has been experimentally observed
\cite{Xu09}, is physically due to the dynamical backaction, i.e.,
the non-Markovian effect, of the memory environments. However, in
many cases the finite extension of the entangled time is not enough
and thus it is desired to preserve a significant fraction of the
entanglement in the long time limit. Indeed, it was shown
\cite{Bellomo08} that some noticeable fraction of entanglement can
be obtained by engineering structured environment such as photonic
band-gap materials \cite{John1990}. According to these works, it is
still unclear if the residual entanglement is fundamentally due to
the specific structured materials or due to certain physical
mechanism. Is there any essential relationship between the ESD
and/or its revival phenomena and the residual entanglement?

In this chapter we focus on these questions and elucidate the
physical nature of the residual entanglement. Before proceeding, it
is helpful to recall the physics of quantum electrodynamics of a
single two-level atom placed in a dielectric with a photonic band
gap \cite{John1994,John1990}. The coupling between the excited atom
and electromagnetic vacuum in the dielectric leads to a novel
photon-atom bound state, in which the fractional atomic population
on the excited state occurs, also known as population trapping
\cite{Lambropoulos00}. This result has been verified experimentally
for quantum dots embedded in photonic band-gap environment
\cite{Lodahl04}. The population trapping has been directly connected
to the entanglement trapping due to the structured environment
\cite{Bellomo08}. Here we reveal for the first time that there are
two essential conditions to preserve the entanglement. One is the
existence of the bound state between the system and its environment,
which provides an ability to preserve the entanglement, and the
other is the non-Markovian effect, which provides a way to preserve
the entanglement. Our result can reproduce the ESD \cite{Yu04} when
the non-Markovian effect is neglected. The phenomenon of the ESD and
its revival discussed in Ref. \cite{Bellomo07} results from the
non-Markovian effect when the bound state is not available. The
interplay between the availability of the bound state and the
non-Markovian effect can lead to a significant fraction of the
entanglement preserved in the steady state. We verify these results
by considering two reservoirs modeled by the super-Ohmic and
Lorentzian spectra, respectively. The result provides a general
method on how to protect the entanglement by engineering the
environment.

\section{The model and entanglement dynamics}

We now consider two spatially separated systems A and B, each has a
two-level atom coupled to a vacuum reservoir, and the two qubits are
initially entangled but have no direct interaction. Owing to the
independence of the two systems \cite{Bellomo07}, we can investigate
single \textquotedblleft qubit + reservior\textquotedblright\ system
at the first place, then extend our studies to the double-one.

The single \textquotedblleft qubit + reservoir\textquotedblright\
system can be formulated by the following Hamiltonian
\begin{equation}
H=\omega _{0}\sigma _{+}\sigma _{-}+\sum_{k}\omega _{k}b_{k}^{\dag
}b_{k}+\sum_{k}(g_{k}\sigma _{+}b_{k}+g_{k}^{\ast }\sigma
_{-}b_{k}^{\dag }), \label{hmtep}
\end{equation}%
where $\omega _{0}$ is the transition frequency of the two-level atom, and $%
\sigma _{\pm }$ are the atom raising and lowering operators,
$b_{k}^{\dag }$ and $b_{k}$ are respectively the creation and
annihilation operators of the \textbf{k}-th mode with frequency
$\omega _{k}$ of the reservoir. $g_k$ denotes the coupling strength
between the atom and the radiation field.

Following the procedure we done in the Section 2.2, we can obtain
the master equation of the qubit,
\begin{eqnarray}
\frac{d\rho ^{S}(t)}{dt} &=&-i\frac{\Omega (t)}{2}[\sigma _{+}\sigma
_{-},\rho ^{S}(t)]\nonumber \\
&&+\frac{\gamma (t)}{2}[2\sigma _{-}\rho ^{S}(t)\sigma _{+}-\sigma
_{+}\sigma _{-}\rho ^{S}(t)-\rho ^{S}(t)\sigma _{+}\sigma _{-}],
\end{eqnarray}%
where $\Omega (t)=-2\text{Im}[\frac{\dot{c}_{0}(t)}{c_{0}(t)}]$, and
$\gamma (t)=-2\text{Re}[\frac{\dot{c}_{0}(t)}{c_{0}(t)}]$. This is
exactly the master equation of the single qubit. $\Omega (t)$ and
$\gamma (t)$ play, respectively, the role of time-dependent Lamb
shift and decay rate.

With the dynamics of single \textquotedblleft qubit + reservoir
\textquotedblright, we can readily study the decoherence dynamics of
the double-one. We assume, for simplicity, the two systems are the
same, and the double-system is
initially in a mixed state%
\begin{equation}
\rho _{tot}^{T}(0)=\left(
\begin{array}{cccc}
\rho _{11} & \rho _{12} & \rho _{13} & \rho _{14} \\
\rho _{12}^{\ast } & \rho _{22} & \rho _{23} & \rho _{24} \\
\rho _{13}^{\ast } & \rho _{23}^{\ast } & \rho _{33} & \rho _{34} \\
\rho _{14}^{\ast } & \rho _{24}^{\ast } & \rho _{34}^{\ast } & \rho _{44}%
\end{array}%
\right) \otimes \left\vert \{0\}_{n}\right\rangle \left\langle
\{0\}_{n}\right\vert .
\end{equation}%
Following the method we applied in the single qubit system, it is
easy to obtain the time evolution of the two qubits.
The diagonal elements are%
\begin{eqnarray}
\rho _{11}^{T}(t) &=&\rho _{11}\left\vert c_{0}(t)\right\vert ^{4},
\notag
\\
\rho _{22}^{T}(t) &=&\rho _{22}\left\vert c_{0}(t)\right\vert
^{2}+\rho _{11}\left\vert c_{0}(t)\right\vert ^{2}(1-\left\vert
c_{0}(t)\right\vert
^{2}),  \notag \\
\rho _{33}^{T}(t) &=&\rho _{33}\left\vert c_{0}(t)\right\vert
^{2}+\rho _{11}\left\vert c_{0}(t)\right\vert ^{2}(1-\left\vert
c_{0}(t)\right\vert
^{2}),  \notag \\
\rho _{44}^{T}(t) &=&1+\rho _{11}\left\vert c_{0}(t)\right\vert
^{4}-\left\vert c_{0}(t)\right\vert ^{2}(2\rho _{11}+\rho _{22}+\rho
_{33}),
\end{eqnarray}%
the nondiagonal elements are
\begin{eqnarray}
\rho _{12}^{T}(t) &=&\rho _{12}\left\vert c_{0}(t)\right\vert
^{2}c_{0}(t),
\notag \\
\rho _{13}^{T}(t) &=&\rho _{13}\left\vert c_{0}(t)\right\vert
^{2}c_{0}(t),
\notag \\
\rho _{14}^{T}(t) &=&\rho _{14}c_{0}^{2}(t),  \notag \\
\rho _{23}^{T}(t) &=&\rho _{23}\left\vert c_{0}(t)\right\vert ^{2},
\notag
\\
\rho _{24}^{T}(t) &=&\rho _{24}c_{0}(t)+\rho
_{13}c_{0}(t)(1-\left\vert
c_{0}(t)\right\vert ^{2}),  \notag \\
\rho _{34}^{T}(t) &=&\rho _{34}c_{0}(t)+\rho
_{12}c_{0}(t)(1-\left\vert c_{0}(t)\right\vert ^{2}),
\end{eqnarray}%
and $\rho _{ij}^{T}(t)=\rho _{ji}^{T\ast }(t).$ Differentiating the
reduced density matrix with respect to time, we can obtain the
equation of motion
for the two qubits%
\begin{eqnarray}
\frac{d\rho ^{T}(t)}{dt} &=&-i\frac{\Omega (t)}{2}([\sigma
_{+}^{A}\sigma
_{-}^{A},\rho ^{T}(t)]+[\sigma _{+}^{B}\sigma _{-}^{B},\rho ^{T}(t)] \nonumber\\
&&+\frac{\gamma (t)}{2}\{[2\sigma _{-}^{A}\rho ^{T}(t)\sigma
_{+}^{A}-\sigma _{+}^{A}\sigma _{-}^{A}\rho ^{T}(t)-\rho
^{T}(t)\sigma _{+}^{A}\sigma
_{-}^{A}] \nonumber\\
&&+[2\sigma _{-}^{B}\rho ^{T}(t)\sigma _{+}^{B}-\sigma
_{+}^{B}\sigma _{-}^{B}\rho ^{T}(t)-\rho ^{T}(t)\sigma
_{+}^{B}\sigma _{-}^{B}]\},
\end{eqnarray}%
where $\Omega (t)$ and $\gamma (t)$ are defined the same as before.

To investigate the entanglement dynamics of the bipartite system, we
apply Wootters concurrence \cite{wootters98}. The concurrence can be
calculated explicitly from the time dependent density matrix $\rho
^{T}(t)$ of the two
qubits, $C(\rho ^{T})=\max \{0,\sqrt{\lambda _{1}}-\sqrt{\lambda _{2}}-%
\sqrt{\lambda _{3}}-\sqrt{\lambda _{4}}\}$, where the quantities
$\lambda
_{i}$ are the eigenvalues of the matrix $\zeta $%
\begin{equation}
\zeta =\rho ^{T}(\sigma _{y}^{A}\otimes \sigma _{y}^{B})\rho ^{T\ast
}(\sigma _{y}^{A}\otimes \sigma _{y}^{B}),
\end{equation}%
arranged in decreasing order. Here $\rho ^{T\ast }$ means the
complex conjugation of $\rho ^{T}$, and $\sigma _{y}$ is the Pauli
matrix. It can be proved that the concurrence varies from 0 for a
separable state to 1 for a maximally entangled state.

As is pointed out, in the Markovian situation the ESD occurs due to
the double excitation of the initial state in a vacuum reservoir
\cite{Ikram07}. In what follows, we consider the initial state of
which the concurrence dynamics can exhibit ESD in Markovian
approximation, and compare this with the non-Markovian situation.

For an initially entangled pure state in the standard bases%
\begin{equation}
\psi(0)=\alpha \left\vert --\right\rangle +\beta \left\vert
++\right\rangle ,  \label{t8}
\end{equation}%
$\alpha $ and $\beta $ satisfy normalization condition. From the
time-dependent reduced density matrix of the two qubits and the
definition
of concurrence, we obtain%
\begin{equation}
C(\rho ^{T})=\max \{0,2\left\vert c_{0}(t)\right\vert ^{2}\left\vert
\beta \right\vert [\left\vert \alpha \right\vert -\left\vert \beta
\right\vert (1-\left\vert c_{0}(t)\right\vert ^{2})]\}.  \label{t12}
\end{equation}%

From the expression of concurrence, it can be found that the
behavior of time-dependent factor local excited state population
($\left\vert c_{0}(t)\right\vert ^{2}$) completely determines the
dynamics of concurrence. In particular, if local
system decoherences completely ($\left\vert c_{0}(\infty)\right\vert ^{2}=0$%
), entanglement would vanish. Whereas, if decoherence in local
system is inhibited, then it is possible to protect entanglement in
the long-time limit.

\section{Mechanism of entanglement preservation}

Entanglement is the fundamental resource of quantum information
processing. Entanglement preservation is a key step for entanglement
applications. To achieve this aim, as we have discussed in Sec. 3.2,
one needs the decoherence suppression in the local system. This
could be obtained when atom-photon bound state is formed.

Following the same procedure done in Sec. 2.4.3, we obtain the
condition for the formation of bound state \cite{tong1,Tong09}
\begin{equation}
 y(E)\equiv \omega _{0}- \int_{0}^{\infty }\frac{J(\omega) }{\omega -
E}d\omega = E, \label{eigen}
\end{equation}
where $J(\omega)=\sum_k |g_k|^2\delta(\omega-\omega_k)$ is the
spectral density of the reservoir. The solution of Eq. (\ref{eigen})
highly depends on the explicit form of the $J(\omega)$. If the
reservoir contains only one mode $\omega^\prime$, then $J(\omega) =
g^2\delta(\omega-\omega^{\prime })$. This is the ideal
Jaynes-Cummings model \cite{Scully97}, in which two bound states in
the one excitation sector are formed and as a result the dynamics of
the system displays a lossless oscillation. When the reservoir
contains infinite modes, one can model $J(\omega)$ by some typical
spectrum functions such as the super-Ohmic or Lorentzian form.

We firstly consider the super-Ohmic spectrum $J(\omega)=\eta
\frac{\omega^3}{\omega_0^2}e^{-\omega/\omega_c}$, where $\eta$ is a
dimensionless coupling constant and $\omega_c$ characterizes the
frequency regime in which the power law is valid \cite{Weiss}. It
corresponds to that the reservoir consists of a vacuum radiation
field, where $g_k\propto \sqrt{\omega_k}$ \cite{Scully97}. The
existence of a bound state requires that Eq. (\ref{eigen}) has at
least a real solution for $E < 0$. It is easy to check that the
solution always exists if the condition $y(0)<0$, i.e. $ \omega
_{0}-2\eta \frac{\omega_c^3}{\omega_0^2}<0$ is satisfied. Otherwise,
no bound state exists. This condition can be fulfilled easily by
engineering the environment. For the Lorentzian spectrum it is found
that a criterion when a bound state exists can not be obtained
analytically. In this case one can use the diagrammatic technique
shown later.

The formation of bound state may lead to the inhibition of
spontaneous emission and results in the population trapping.
Therefore the formation of local bound state has a profound effect
on the preservation of non-local coherence(entanglement). One could
imagine that non-Markovian effect is also a key factor to preserve
entanglement since non-Markovian effect is a kind of backaction
effect which could compensate the lost coherence of quantum system.
In the following, we explicitly study the formation of bound state
and non-Markovian effect on entanglement dynamics using the examples
of super-ohmic and Lorentzian vacuum reservoir \cite{Tong09}.

\begin{figure}[tbp]
\includegraphics[width = \columnwidth] {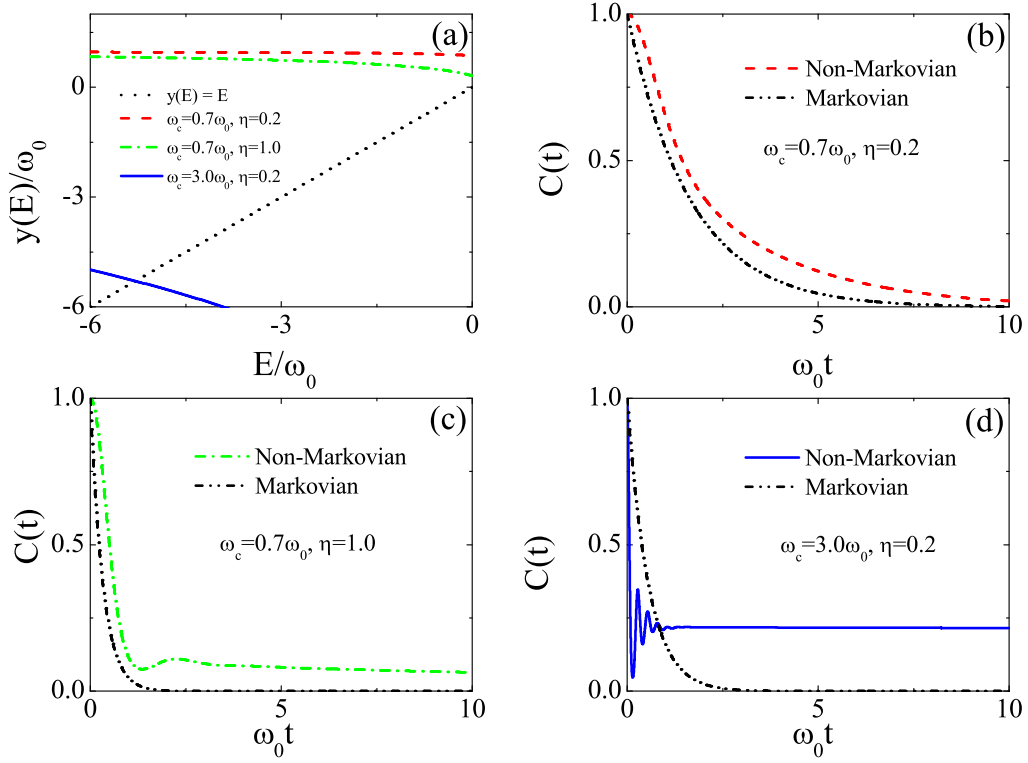}
\caption{ Entanglement dynamics of the two-qubit system with local
super-Ohmic reservoirs. (a) Diagrammatic solutions of Eq.
(\ref{eigen}) with different parameters. $C(t)$ as a function of
time is shown in (b): $(\omega_c, \eta) = (0.7\omega_0, 0.2)$, (c):
$(\omega_c, \eta) = (0.7\omega_0, 1.0)$ and (d): $(\omega_c, \eta) =
(3.0\omega_0, 0.2)$. The parameter $\alpha$ is taken as $0.7$. For
comparison, $C(t)$ under the Markovian approximation has also been
presented by using the same parameters. } \label{fig1}
\end{figure}
\begin{figure}[h]
\begin{center}
\includegraphics[width = 0.8\columnwidth]{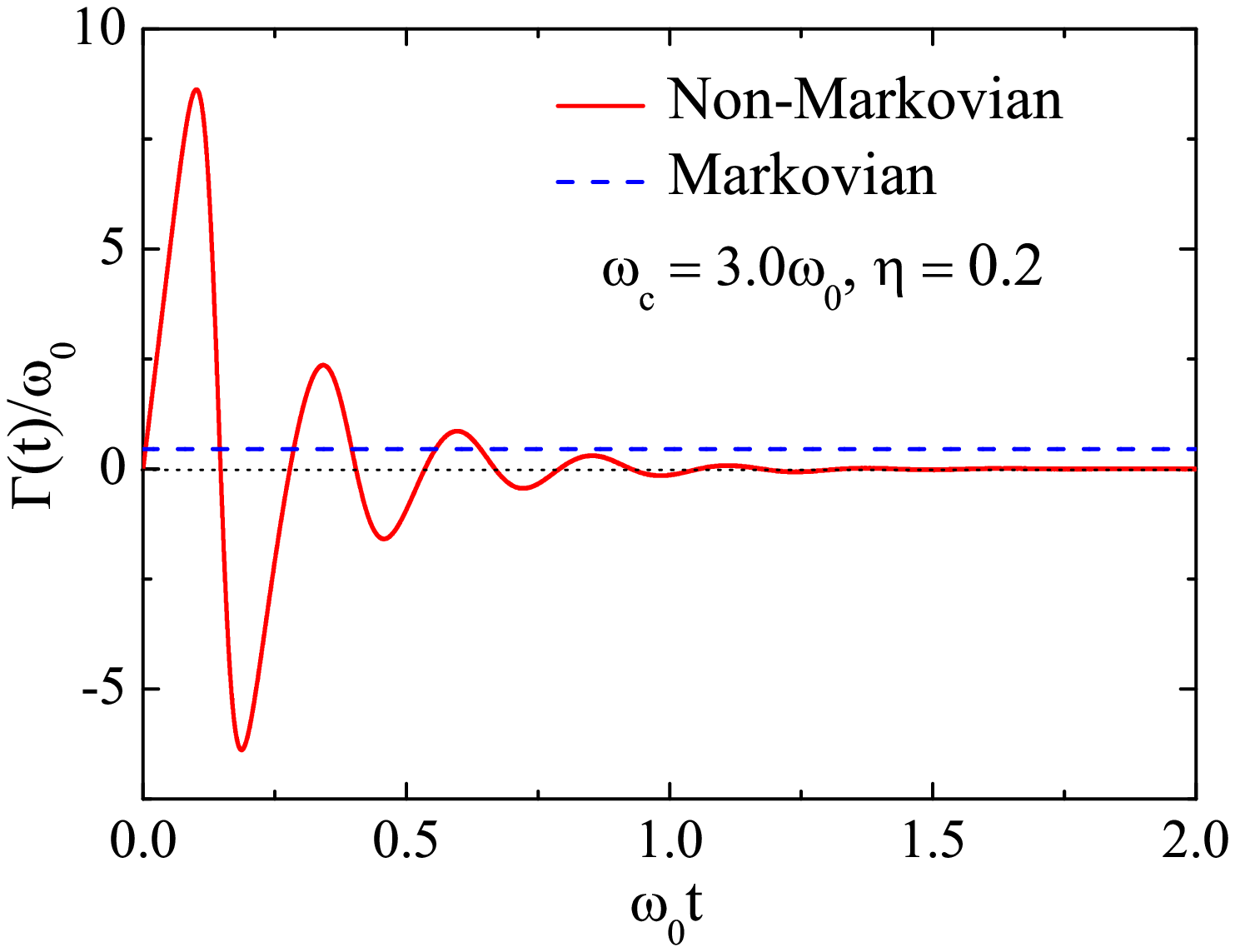}
\caption{ The decay rate $\Gamma(t)$ as a function of time in the
non-Markovian and Markovian cases. The parameters used are $\omega_c
= 3.0\omega_0$ and $\eta = 0.2$.}  \label{fig2}
\end{center}
\end{figure}

Consider firstly the super-Ohmic case. Fig. \ref{fig1} shows the
entanglement dynamics in different parameter regimes, i.e.,
$(\omega_c, \eta) = (0.7\omega_0, 0.2), (0.7\omega_0, 1.0)$ and
$(3.0\omega_0, 0.2)$. For the first two parameter sets the bound
state is absent, while for the last one it is available, as shown in
Fig. \ref{fig1} (a). Whether the bound state exists or not plays a
key role in the entanglement preservation in the long time limit.
When the bound state is absent, the residual entanglement approaches
zero in a long enough time, as shown by the solid lines in Fig.
\ref{fig1} (b) and (c). Difference between these two cases is that
Fig. \ref{fig1}(b) is in weak coupling regime, where the
non-Markovian effect is weak, while Fig. \ref{fig1}(c) is in strong
coupling regime, where the strong non-Markovian effect leads to an
obvious oscillation. When the bound state is available, the
situation is quite different, as shown in Fig. \ref{fig1}(d). The
entanglement firstly experiences some oscillations due to the energy
and/or information exchanging back and forth between the qubit and
its memory environment \cite{Pillo08}, then approaches a definite
value in the long time limit, where the decay rate approaches zero
after some oscillations, as shown in Fig. \ref{fig2}. {\it The
entanglement preservation is a result from the interplay between the
existence of the bound state (providing an ability to preserve the
entanglement) and the non-Markovian effect (providing a way to
preserve the entanglement)} \cite{Tong09}. The claim can be further
verified by that the entanglement preservation is absent in the
Markovian dynamics, as shown by the dashed lines in Fig. \ref{fig1}
(b), (c) and (d), where the entanglement displays sudden death
irrespective of the availability of the bound state. It is because
the Markovian environment has no memory and the energy/information
flowing from the qubit to its environment is irreversible and the
decay rate keeps to be fixed (see, Fig. \ref{fig2}). In this case
one has
\begin{equation}
C(t)=\max \{0,2e^{-2\Gamma_0t}\left\vert \beta \right\vert
[\left\vert \alpha \right\vert -\left\vert \beta \right\vert
(1-e^{-2\Gamma_0t})]\}, \label{cMar}
\end{equation}
which shows a finite disentanglement time when $|\alpha|<|\beta|$
\cite{Yu04}. In a word, the above discussion manifests clearly two
conditions to preserve the entanglement, i.e., the availability of
the bound state and the non-Markovian effect \cite{Tong09}, not only
the structured environment as emphasized in Ref. \cite{Bellomo08}.

\begin{figure}[tbp]\begin{center}
\includegraphics[width = 0.8\columnwidth]{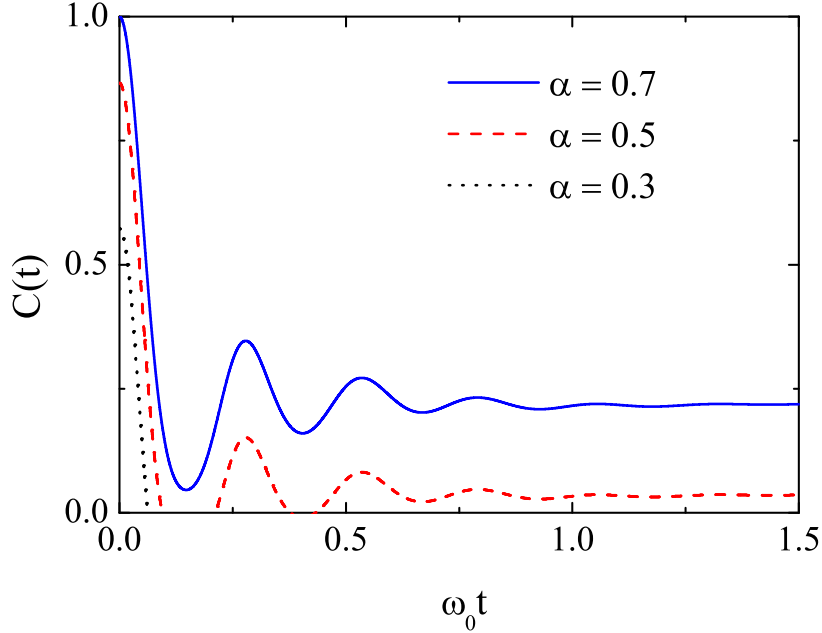}
\caption{ The residual entanglement for different initial states
with $\alpha = 0.7, 0.5$ and $0.3$. The other parameters used are
the same as those in Fig. \ref{fig1}. } \label{fg}\end{center}
\end{figure}

The above discussion focused on almost maximally entangled initial
state by taking $\alpha = 0.7$. In Fig. \ref{fg} we show the results
for different initial states with different initial entanglement.
With decreasing the initial entanglement, the residual entanglement
also decreases in the long time limit and finally, the ESD happens
for $\alpha = 0.3$. The result can be understood from Eq.
(\ref{t12}). On the one hand, the residual entanglement is
determined by $c_0(\infty)$, which is directly related to the
property of the bound state. On the other hand, the residual
entanglement is also determined by the competition between the first
and the second terms in Eq. (\ref{t12}), which is dependent of the
initial state.

In order to make a comparative study and confirm our observations we
consider the Lorentzian spectrum if the reservoir is composed of
lossy cavity,
\begin{equation}
J(\omega )=\frac{1}{2\pi }\frac{\gamma \lambda ^{2}}{(\omega
-\omega_0 )^{2}+\lambda ^{2}},  \label{t11}
\end{equation}%
where $\gamma $ is the coupling constant and $\lambda $ is the
spectrum width. This model has also been studied in Ref.
\cite{Bellomo07}, where the lower limit of frequency integral in
$f(t-\tau)$ was extended from zero to negative infinity. This
extension is mathematically convenient but the availability of the
bound state is missed. Here we follow the original definition of the
frequency integral ranges.

\begin{figure}[tbp]
\includegraphics[width = \columnwidth] {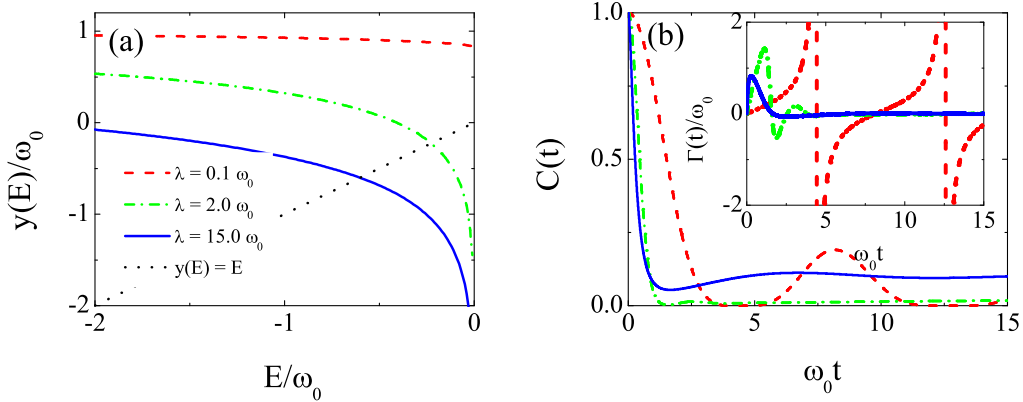}
\caption{ The entanglement dynamics with the Lorentzian spectrum.
(a) Diagrammatic solutions of Eq. (\ref{eigen}) with different
parameters $\lambda = 0.1 \omega_0, 2.0\omega_0$ and $15\omega_0$.
(b) $C(t)$ as a function of time for the corresponding three
parameter regimes. The insert in (b) are the decay rate as a
function of time. The other parameters used are $\gamma =
3.0\omega_0$ and $\alpha = 0.7$. } \label{fig3}
\end{figure}

\begin{figure}[h]
\includegraphics[width = \columnwidth] {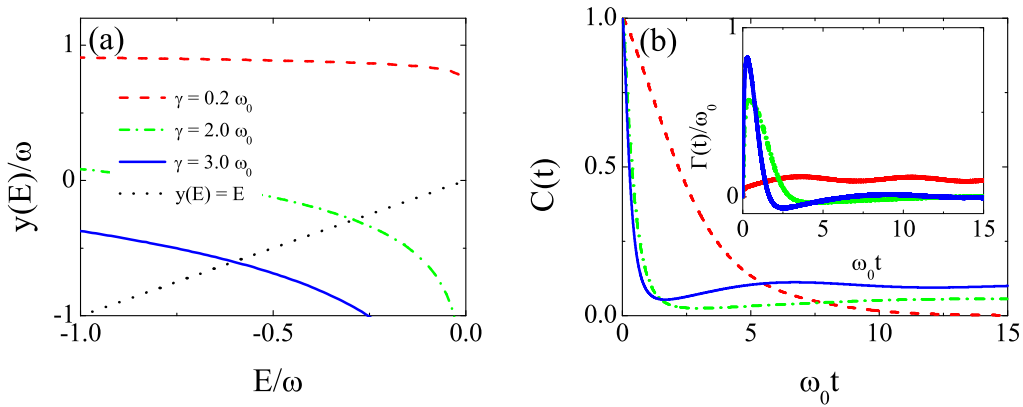}
\caption{The same as Fig. \ref{fig3} but $\lambda = 15.0\omega_0$ is
fixed and $\gamma = 0.2\omega_0, 2.0\omega_0$ and $3.0\omega_0$.}
\label{fig4}
\end{figure}

Our model with the Lorentzian spectral density corresponds exactly
to the extended damping J-C model \cite{Breuer02}. It is noted that
the strong coupling of J-C model has been achieved in circuit QED
\cite{Wallraff04} and quantum dots \cite{Hennessy07} systems. Fig.
\ref{fig3} shows the entanglement dynamics of the qubits under the
Lorentizan reservoir for different spectrum widths in the strong
coupling regime. When $\lambda = 0.1\omega_0$, Eq. (\ref{eigen})
lacks the bound state. According to the above discussion, there is
no residual entanglement in the long time limit. This is indeed
true, as shown in Fig. \ref{fig3} (b). However, it is noted that
before becoming zero the entanglement exhibits ``sudden death" and
revives after some time for several times. This is an analog of the
central result found in Ref. \cite{Bellomo07}, i.e., the phenomenon
of the ESD and revival. Apparently, this is due to the non-Markovian
effect, the revival is a result of backaction of the memory
reservoir. When increasing the spectrum width the bound states
become available, the situation changes. The significant fraction of
the entanglement initially present is preserved in the long time
limit, where the decay rates shown in the insert of Fig. \ref{fig3}
(b) approach zero in these cases. Likewise, the physical nature of
the entanglement preservation is still the interplay between the
bound state and the non-Markovian effect. The more stronger the
coupling is, the more striking the entanglement oscillates as a
function of time, consequently, the more noticeable the
non-Markovian effect is, as shown in Fig. \ref{fig4}. For $\gamma =
0.2 \omega_0$, the system is in the weak coupling regime, where the
bound state is also not available. As a result, the ESD is
reproduced in this case.
\begin{figure}[h]\begin{center}
\includegraphics[width = 0.8\columnwidth]{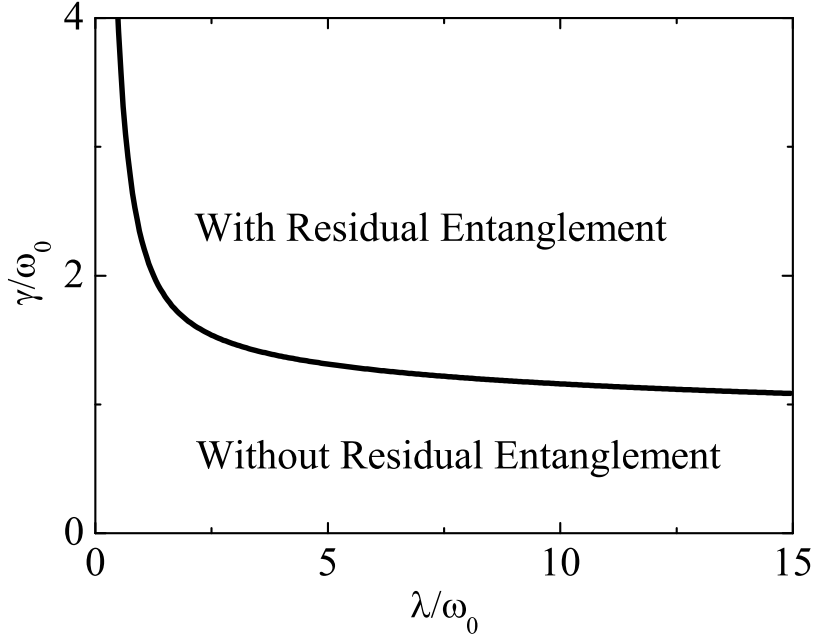}
\caption{The phase diagram of the residual entanglement in the
steady state for Lorentzian spectrum.} \label{fig5}\end{center}
\end{figure}

In Fig. \ref{fig5} we present a phase diagram of the entanglement in
the steady state for the Lorentzian spectral density. In the large
$\gamma$ and small $\lambda$ regime, the system approaches the J-C
model. In this situation the strong backaction effect of the
reservoirs makes the qubit system hard to form a steady state. The
entanglement oscillates with time but has no dissipation. In the
small $\gamma$ and large $\lambda$ regime, the non-Markovian effect
is extremely weak and our results reduce to the Markovian case. In a
limit of the flat spectral density, the Born-Markovian approximation
is applicable and the system has no bound state. This is the case of
the ESD \cite{Yu04,Tong09}.

\section{Summary}\label{cd}
In this chapter, we have studied the entanglement protection of two
qubits in two uncorrelated reservoirs. Two essential conditions to
preserve the entanglement are explored, one is the existence of the
bound state of the system and its reservoir and the other one is the
non-Markovian effect. The bound state provides the ability of the
entanglement preservation and the non-Markovian effect provides the
way to protect the entanglement. The previous results on the
entanglement dynamics in the literature can be considered as the
specific cases where these two conditions have not been fulfilled at
the same time. The result provides a unified picture for the
entanglement dynamics and gives a clear way how to protect the
entanglement. This is quite significant in the quantum information
processing.

The presence of such entanglement preservation gives us an active
way to suppress decoherence. This could be achieved by modifying the
spectrum of the reservoirs to approach the non-Markovian regime and
form a bound state via the potential usage of the reservoir
engineering technique \cite{Wineland001,Turchette00,Garraway06}.
Fortunately, we notice that many practical systems have now be
engineered to show strong non-Markovian effect
\cite{Dubin07,Koppens07,Mogilevtsev08,Xu09}. All these achievements
have paved the way to experimentally simulate the paradigmatic
models of open quantum system, which gives a hopeful prospective to
preserve the entanglement.

\chapter{Entanglement distribution and its invariance}\label{edvtq}

In this chapter we study the entanglement distribution among
bipartite systems of quantum systems and their reservoirs. Following
the method used in Ref. \cite{Lopez08}, we find an invariant and
entanglement can be distributed among all the bipartite subsystems.

\section{Introduction}

In Chapter 3, we have study the entanglement dynamics under
different environments. We have figured out the mechanism of
entanglement preservation, i.e. entanglement can be preserved when
bound states are formed under the non-Markovian dynamics.

Recently, L\'{o}pez \textit{et al.} asked a question about where the
lost entanglement between the qubits goes \cite{Lopez08}.
Interestingly, they found that the lost entanglement of the qubits
is exclusively transferred to the reservoirs under the Markovian
dynamics and the ESD of the qubits is always accompanied with the
entanglement sudden birth (ESB) of the reservoirs. This means that
the entanglement does not go away, it is still there but just
changes the location. This is reminiscent of the work of Yonac
\textit{et al.} \cite{Yonac}, in which the entanglement dynamics has
been studied in a double J-C model. They found that the entanglement
is transferred periodically among all the bipartite partitions of
the whole system but an identity (see below) has been satisfied at
any time. This may be not surprising since the double J-C model has
no decoherence and any initial information can be preserved in the
time evolution. However, it would be surprising if the identity is
still valid in the presence of the decoherence, in which a
non-equilibrium relaxation process is involved. In this chapter, we
show that it is indeed true for such a system consisted of two
qubits locally interacting with two reservoirs. We find that the
distribution of the entanglement among the bipartite subsystems is
dependent of the explicit property of the environment and its
coupling with the qubit. The rich dynamical behaviors obtained
previously in the literature can be regarded as the special cases of
our present result or Markovian approximation. Particularly, we find
that the entanglement can stably distribute among all the bipartite
subsystems if the qubit and its environment can form a bound state
and the non-Markovian effect is important. Irrespective of how
distributes the entanglement, it is found that the identity about
the entanglement in the whole system can be satisfied at any time,
which reveals the profound physics of the entanglement dynamics.

\section{The model of two qubits in two uncorrelated band-gap reservoirs}\label{med}

The model we studied here is the same as the one used in chapter 3,
i.e. two qubits in two separate vacuum reservoirs. Because of the
independence of the two subsystems, we first consider the single
``qubit + reservoir'' subsystem which is governed by the following
Hamiltonian
\cite{Scully97}%
\begin{equation}
H=\omega _{0}\sigma _{+}\sigma _{-}+\sum_{k}\omega _{k}a_{k}^{\dag
}a_{k}+\sum_{k}(g_{k}\sigma _{+}a_{k}+h.c.),  \label{t1}
\end{equation}%
in which the notations are the same as Sec. 2.2

As we have discussed in chapter 2, the initial state of the local
system $\left\vert \phi (0)\right\rangle =\left\vert -\right\rangle
\otimes \left\vert \{0\}_{k}\right\rangle $, where $\left\vert
\{0\}_{k}\right\rangle $ denotes the vacuum state of reservoir, does
not evolve with time. While for an initial state $\left\vert \phi
(0)\right\rangle =\left\vert +\right\rangle
\otimes \left\vert \{0\}_{k}\right\rangle $, governed by the Hamiltonian (%
\ref{t1}), its time evolution is given by%
\begin{equation}
\left\vert \phi (t)\right\rangle =b(t)\left\vert
+,\{0\}_{k}\right\rangle +\sum_{k}b_{k}(t)\left\vert
-,\{1\}_{k}\right\rangle ,  \label{t9}
\end{equation}%
where $\left\vert \{1\}_{k}\right\rangle $ denotes the reservoir
state with only one photon in the k-th mode. From the
Schr\"{o}dinger equation, we can get the time evolution of the
excited probability amplitudes in Eq. (\ref{t9})
\begin{equation}
\dot{b}(t)+i\omega _{0}b(t)+\int_{0}^{t}b(\tau )f(t-\tau )d\tau=0 ,
\label{t7}
\end{equation}%
where the kernel function is $f(x)=\int_{0}^{\infty }d\omega
J(\omega )e^{-i\omega x}$ with $J(\omega )=\sum_{k}\left\vert
g_{k}\right\vert ^{2}\delta (\omega -\omega _{k})$ being the
spectral density.

If we define the normalized collective state with one excitation in
the reservoir as $\left\vert \mathbf{\tilde{1}}\right\rangle
_{r}=[\sum_{k}b_{k}(t)\left\vert \{1\}_{k}\right\rangle
]/\tilde{b}(t)$ and with no excitation in the reservoir as
$\left\vert \mathbf{\tilde{0}}\right\rangle _{r}=\left\vert
\{0\}_{k}\right\rangle $, then Eq. (\ref{t9}) can be recast into
\cite{Lopez08,tong3}
\begin{equation}
\left\vert \phi (t)\right\rangle =b(t)\left\vert +\right\rangle
\left\vert \mathbf{\tilde{0}}\right\rangle
_{r}+\tilde{b}(t)\left\vert -\right\rangle \left\vert
\mathbf{\tilde{1}}\right\rangle _{r},  \label{t10}
\end{equation}%
where $\tilde{b}(t)=\sqrt{1-\left\vert b(t)\right\vert ^{2}}$.

According to the above results, the time evolution of a system
consisted of two such subsystems with the initial state $\left\vert
\Phi (0)\right\rangle =(\alpha \left\vert -,-\right\rangle +\beta
\left\vert +,+\right\rangle )\left\vert
\mathbf{\tilde{0}}\right\rangle _{r_{1}}\left\vert
\mathbf{\tilde{0}}\right\rangle _{r_{2}}$ is given by
\begin{equation}
\left\vert \Phi (t)\right\rangle =\alpha \left\vert -,-\right\rangle
\left\vert \mathbf{\tilde{0}}\right\rangle _{r_{1}}\left\vert \mathbf{\tilde{%
0}}\right\rangle _{r_{2}}+\beta \left\vert \phi (t)\right\rangle
_{1}\left\vert \phi (t)\right\rangle _{2}, \label{phit}
\end{equation}
where $\alpha$ and $\beta$ are the coefficients to determine the
initial entanglement in the system. From
$\rho=|\Phi(t)\rangle\langle\Phi(t)|$, one can obtain the
time-dependent reduced density matrix of the bipartite subsystem
qubit1-qubit2 ($q_1q_2$) by tracing over the reservoir variables. It
reads
\begin{equation}
\rho _{q_1q_2}(t)=\left(
\begin{array}{cccc}
|\beta |^{2}\left\vert b(t)\right\vert ^{4} & 0 & 0 & \beta \alpha
^{\ast
}b(t)^{2} \\
0 & p & 0 & 0 \\
0 & 0 & p & 0 \\
\beta ^{\ast }\alpha b^*(t)^{2} & 0 & 0 & x
\end{array}
\right) ,
\end{equation}
where $p=\left\vert \beta b(t)\right|^2\tilde{b}(t)^2$ and
$x=1-|\beta |^{2}|b(t)|^{4}-2p$. By the similar procedure, it is not
difficult to obtain the corresponding reduced density matrices for
other bipartite subsystems like reservoir1-reservoir2 ($r_1r_2$) and
qubit-reservoir ($q_1r_1$, $q_1r_2$, $q_2r_1$, $q_2r_2$).

\section{Entanglement distribution among bipartite subsystems}

Follow chapter 3, we use the concurrence \cite{wootters98} to
quantify entanglement. The concurrence for each bipartite partition
can be calculated as $C_{m}=\max \{0,Q_{m}\}$, where $m$ denotes the
different bipartite partitions and $Q_{m}$ read \cite{tong3}
\begin{eqnarray}
&&Q_{q_{1}q_{2}} = 2|\alpha\beta||b(t)|^2 - 2|\beta b(t)|^2\tilde{b}(t)^{2},  \label{qq} \\
&&Q_{r_{1}r_{2}} = 2|\alpha\beta|\tilde{b}(t)^2 - 2|\beta b(t)|^2\tilde{b}(t)^{2},  \label{rr} \\
&&Q_{q_{i}r_{i}} = 2|\beta|^2|b(t)|\tilde{b}(t) \,\, (i = 1, 2), \label{qr1} \\
&&Q_{q_{1}r_{2}} = 2|\alpha\beta b(t)|\tilde{b}(t) - 2|\beta
b(t)|^2\tilde{b}(t)^{2} = Q_{q_{2}r_{1}}. \label{qr2}
\end{eqnarray}
It is straightforward to verify that the quantities $Q_m$ in Eqs.
(\ref{qq})-(\ref{qr2}) satisfy an identity
\begin{equation}
Q_{q_{1}q_{2}}+Q_{r_{1}r_{2}}+2\left\vert
\frac{\alpha}{\beta}\right\vert
Q_{q_{1}r_{1}}-2Q_{q_{1}r_{2}}=2\left\vert \alpha \beta \right\vert,
\label{t15}
\end{equation}
where $2|\alpha\beta|$ is just the initial entanglement. Eq.
(\ref{t15}) has been obtained in a double J-C model \cite{Yonac}, in
which the decoherence is absent since each of the reservoirs only
contains one mode, i.e. $J(\omega )=g^{2}\delta (\omega -\omega
_{0})$. Surprisingly, this identity is still true even in the
presence of decoherence. Furthermore, one notes that the identity is
not dependent of any detail about $b(t)$, which is determined by Eq.
(\ref{t7}). This result shows clearly the invariant nature of the
entanglement. In the following we explicitly discuss the
distribution behavior of the entanglement by taking the reservoir as
a photonic band gap (PBG) medium \cite{Yablonovitch87,Lodahl04} and
compare it with the previous results.

For the PBG medium, the dispersion relation near the upper band-edge
is given by \cite{John1994}
\begin{equation}
\omega _{k}=\omega_{c}+A(k-k_{0})^{2}, \label{dispersion}
\end{equation}
where $A\approx \omega_{c}/k_{0}^{2}$, $\omega _{c}$ is the upper
band-edge frequency and $k_{0}$ is the corresponding characteristic
wave vector. In this case, the kernel function has the form
\cite{tong3}
\begin{equation}
f(t-\tau)=\eta\int \frac{c^3k^{2}}{\omega_k}e^{-i\omega_k (t-\tau)}
dk,  \label{kn}
\end{equation}
where $\eta = \frac{\omega _{0}^{2}d^{2}}{6\pi^{2}\varepsilon
_{0}c^3 }$ is a dimensionless constant. In solving Eq. (\ref{t7})
for $b(t)$, Eq. (\ref{kn}) is evaluated numerically. Here we do not
make an assumption that $k$ can be replaced by $k_{0}$ outside of
the exponential \cite{John99}, as also done in Refs.
\cite{Bellomo08,Li09,Wang08}. Thus our result is numerically exact.
In the following discussion we take $\omega_c$ as the unit of
frequency.

\begin{figure}[tbp]
\begin{center}
\begin{tabular}{cc}
\resizebox{60mm}{!}{\includegraphics{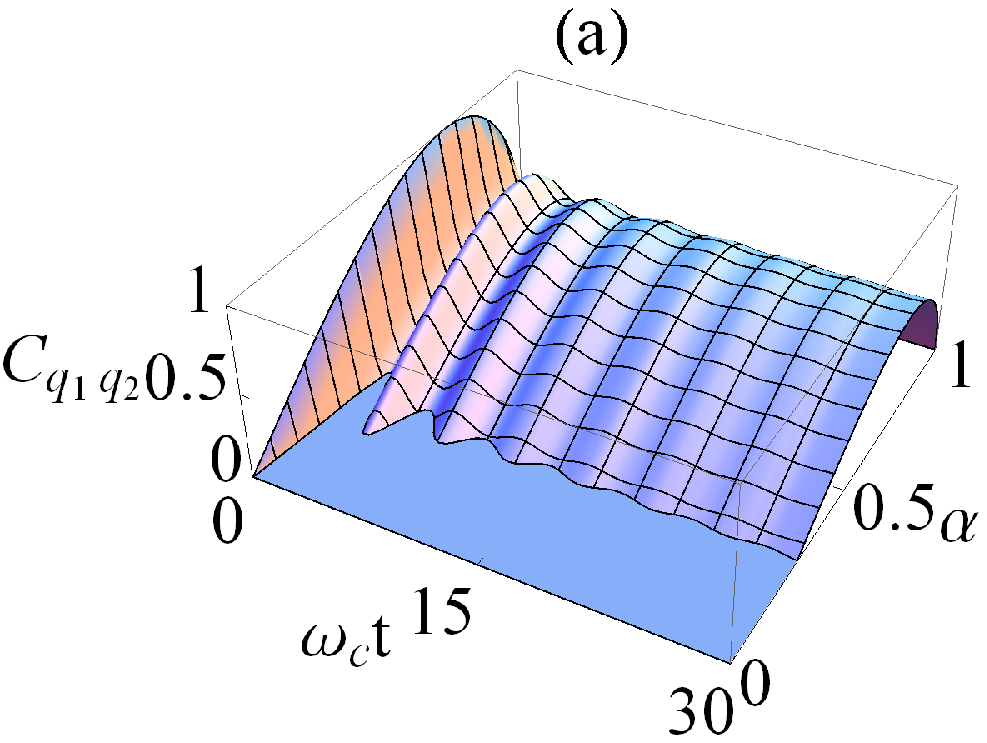}} &
\resizebox{60mm}{!}{
\includegraphics{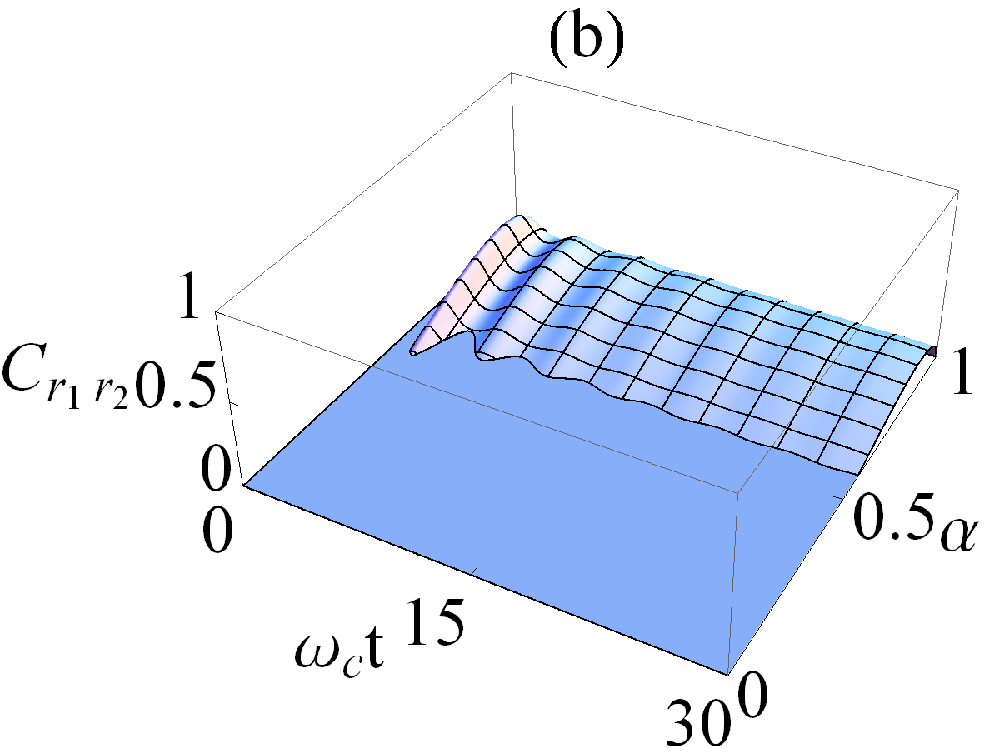}} \\
\resizebox{60mm}{!}{\includegraphics{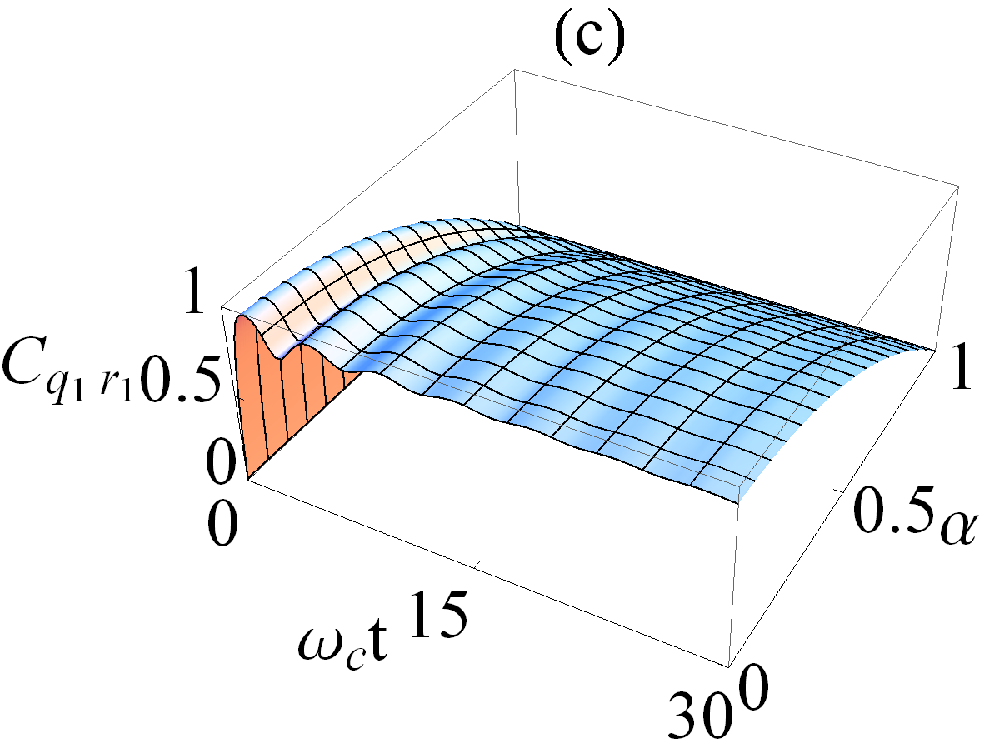}} &
\resizebox{60mm}{!}{
\includegraphics{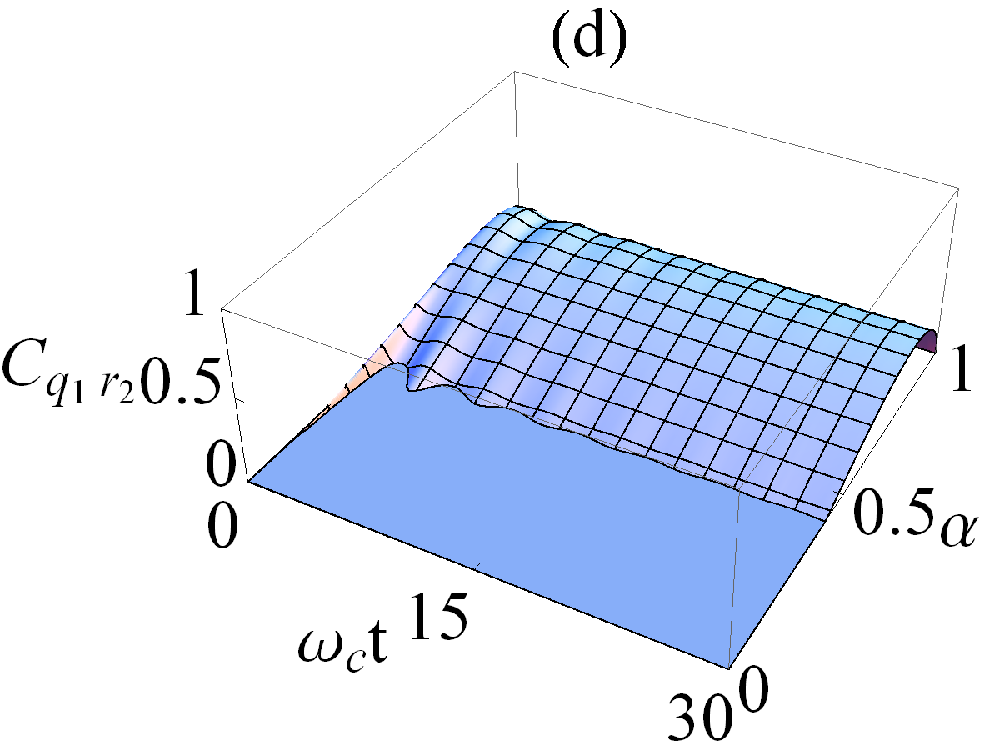}} \\
&
\end{tabular}
\end{center}
\caption{Entanglement distributions and their time evolutions for
the case of $\omega_0 < \omega_c$. The parameters used are
$\omega_{0}=0.1\omega_c$ and $\eta = 0.2$.}\label{bst}
\end{figure}

\begin{figure}[h]
\begin{center}
\begin{tabular}{cc}
\resizebox{60mm}{!}{\includegraphics{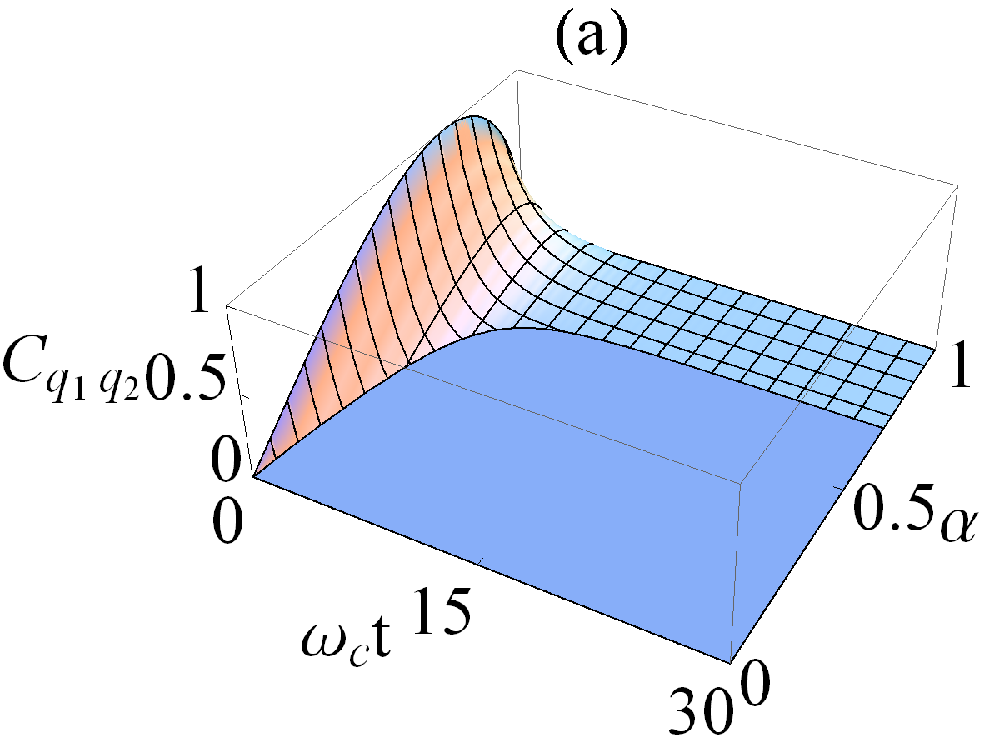}} &
\resizebox{60mm}{!}{
\includegraphics{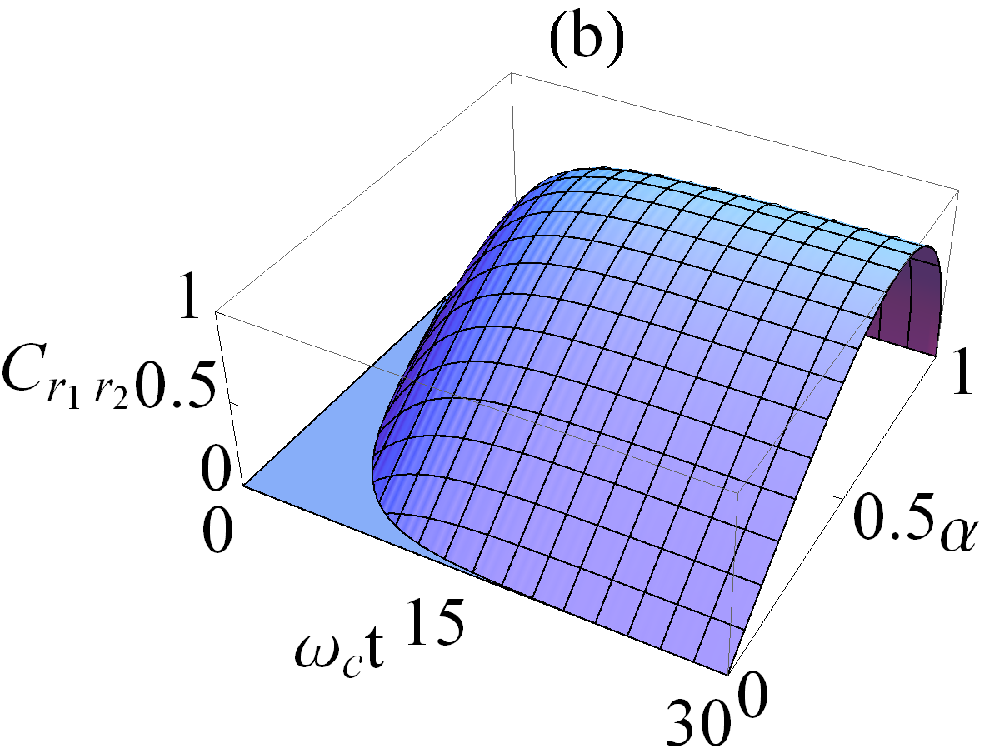}} \\
\resizebox{60mm}{!}{\includegraphics{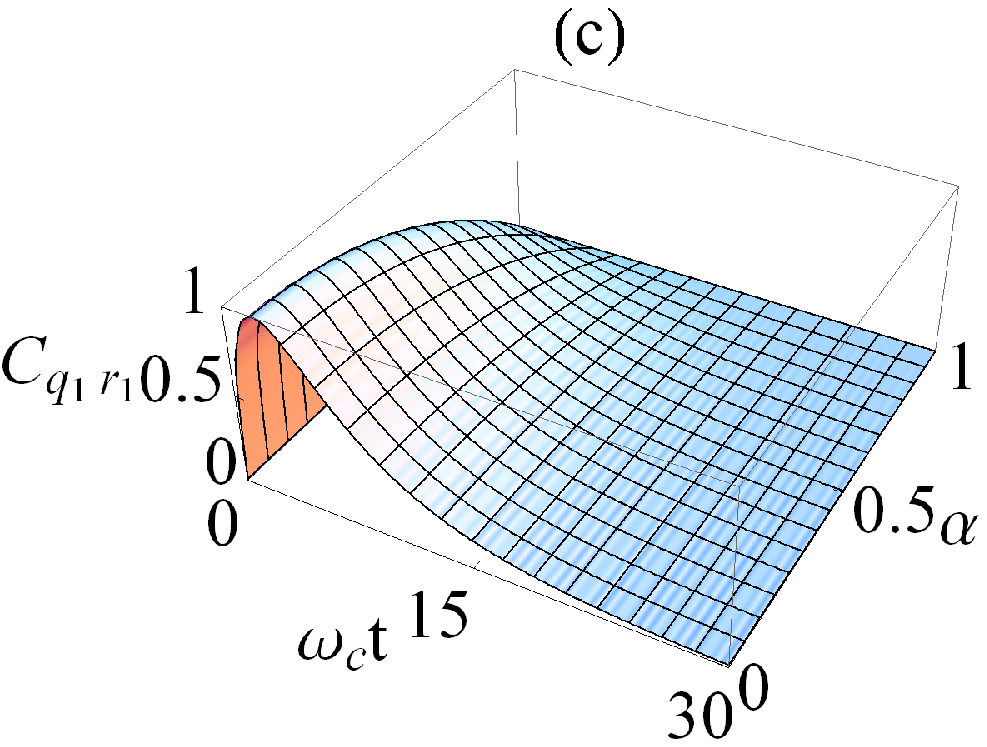}} &
\resizebox{60mm}{!}{
\includegraphics{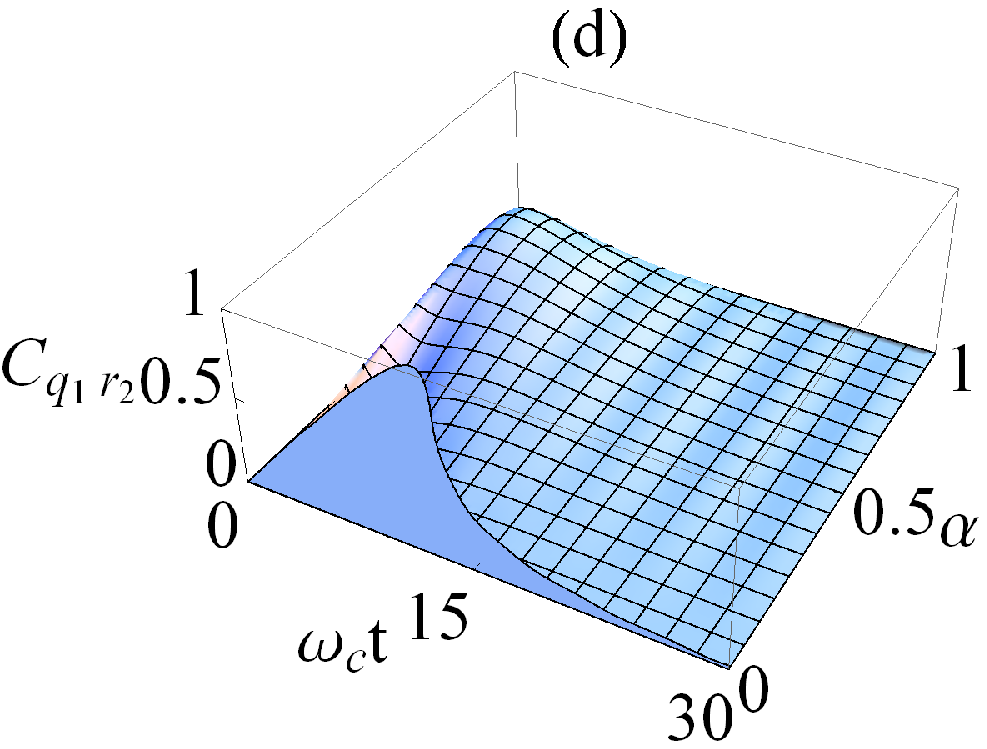}} \\
&
\end{tabular}
\end{center}
\caption{Entanglement distributions and their time evolutions for
the case of $\omega_0 > \omega_c$.  The parameters used are $\omega
_{0}=10.0\omega_c$ and $\eta = 0.2$.} \label{nbst}
\end{figure}

\begin{figure}[h]
\begin{center}
\includegraphics[width = 0.8 \columnwidth]{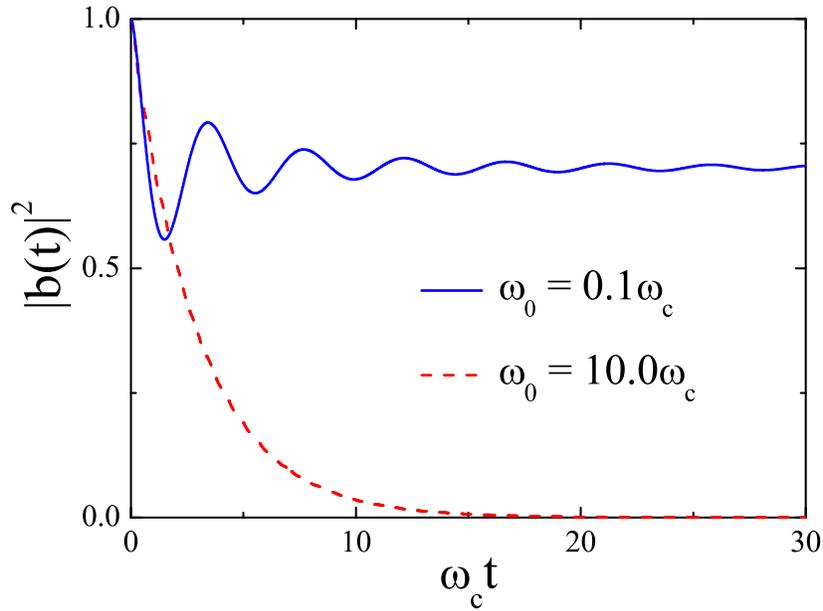}
\end{center}
\caption{Time evolution of time-dependent factor of the
excited-state population for two parameter regimes $\omega_0 =
0.1\omega_c$ (solid line) and $10.0\omega_c$ (dashed line). $\eta$
is taken as 0.2. } \label{tong0}
\end{figure}

In Figs. \ref{bst} and \ref{nbst}, we show the entanglement
distributions and their time evolutions for two typical cases of
$\omega_0 < \omega_c$ and $\omega_0
> \omega_c$, which correspond to the atomic frequency being located at the band
gap and at the upper band of the PBG medium, respectively
\cite{tong3}. In the both cases the initial entanglement in $q_1q_2$
begins to transfer to other bipartite partitions with time but their
explicit evolutions, in particular the long time behaviors, are
quite different. In the former case, the entanglement could
distribute stably among all bipartite partitions. Fig. \ref{bst}(a)
shows that after some oscillations, a sizeable entanglement of
$q_1q_2$ is preserved for the parameter regime of $0.3 < \alpha <1 $
in the long-time limit. Remarkably, the entanglement in
$q_ir_i(i=1,2)$ forms quickly in the full range of $\alpha$ [Fig.
\ref{bst}(c)] and dominates the distribution. On the contrary, only
slight entanglement of $r_1r_2$ is formed in a very narrow parameter
regime $0.6< \alpha <1 $, as shown in Fig. \ref{bst}(b). However,
when $\omega_0$ is located at the upper band of the PBG medium, the
initial entanglement in $q_1q_2$ is transferred completely to the
$r_1r_2$ in the long-time limit, as shown in Fig. \ref{nbst}. At the
initial stage, $q_ir_i (i=1,2)$ and $q_1r_2 (q_2r_1)$ are entangled
transiently, but there is no stable entanglement distribution. This
result is consistent with that in Refs.
\cite{Lopez08,Zhou09,XuZY09}. It is not difficult to understand
these results according to Eqs. (\ref{qq})-(\ref{qr2}). From these
equations, one can clearly see that the detailed behavior of the
entanglement dynamics and its distributions in the bipartite
partitions are completely determined by the time-dependent factor
$|b(t)|^{2}$ of single-qubit excited-state population. Fig.
\ref{tong0} shows its time evolutions for the corresponding
parameter regimes presented above. We notice that $|b(\infty)|^2\neq
0$ when $\omega_0$ is located at the band gap, which means that
there is some excited-state population in the long-time limit. This
is just the population trapping which we have discussed in above
chapters. Such population trapping just manifests the formation of
bound states between $q_i$ and $r_i$ \cite{Tong09}. Consequently,
$q_i$ and $r_i$ are so correlated in the bound states that the
initial entanglement in $q_1q_2$ cannot be fully transferred to
$r_1r_2$. The oscillation during the evolution is just the
manifestation of the strong non-Markovian effect induced by the
reservoirs. On the contrary, if $\omega_0$ is located in the upper
band, then $|b(\infty)|^2=0$ and the qubits decay completely to
their ground states. In this case the bound states between $q_i$ and
$r_i$ are absent and the initial entanglement in $q_1q_2$ is
completely transferred to the $r_1r_2$, as clearly shown in Eq.
(\ref{rr}).

\begin{figure}[tbp]
\begin{center}
\includegraphics[width = \columnwidth]{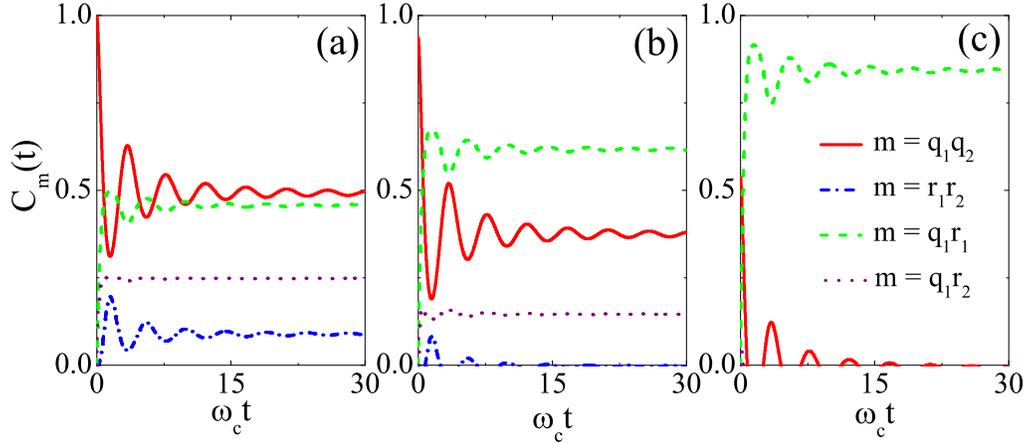}
\end{center}
\caption{(Color online) Entanglement evolution when $\alpha=1/%
\sqrt{2}$ (a), $\alpha=0.57$ (b), and $\protect\alpha=0.28$ (c). The
parameters used here are the same as Fig. \ref{bst}.} \label{td}
\end{figure}

In addition, in Refs. \cite{Lopez08,Zhou09,XuZY09} it was emphasized
that the ESD of $q_1q_2$ is always accompanied with the ESB of
$r_1r_2$. However, this is not always true. To clarify this, we
examine the condition to obtain the ESD of the qubits and the
companying ESB of the reservoirs. From Eqs. (\ref{qq}) and
(\ref{rr}) it is obvious that the condition is $Q_{q_{1}q_{2}}(t)<0$
and $ Q_{r_{1}r_{2}}(t^{\prime })>0$ at any $t$ and $t^{\prime }$,
which means \cite{tong3}
\begin{equation}
\left\vert b(t^{\prime })\right\vert ^{2}<| \alpha |
/\sqrt{1-|\alpha | ^{2}} <1-\left\vert b(t)\right\vert ^{2}.
\label{dd}
\end{equation}
In the case without bound states, $\left\vert b(\infty )\right\vert
^{2}=0$. The condition (\ref{dd}) can be satisfied when
$\alpha<1/\sqrt{2}$. So one can always expect the ESD of the qubits
and the companying ESB of the reservoirs in the region $\left\vert
\alpha \right\vert <1/\sqrt{2}$, as shown in Fig. \ref{nbst} and
Refs. \cite{Lopez08,Zhou09,XuZY09}. However, when the bound states
are available, the situation changes. In particular, when
$\left\vert b(t)\right\vert ^{2}\geq \frac{1}{2}$ in the full range
of time evolution, no region of $\alpha $ can make the condition
(\ref{dd}) to be satisfied anymore. For clarification, we present
three typical behaviors of the entanglement distribution in Fig.
\ref{td}. In all these cases the bound states are available. Fig.
\ref{td} (a) shows the situation where the entanglement is stably
distributed among all of the bipartite subsystems. In Fig. \ref{td}
(b) the entanglement of $ r_1r_2$ shows ESB and revival. However,
the entanglement in $q_1q_2$ does not exhibit ESD. This is the
example that the ESD in $q_1q_2$ is not accompanied with the ESB in
$r_1r_2$. Fig. \ref{td}(c) shows another example that while the
entanglement of $q_1q_2$ shows ESD and revival \cite{Bellomo07}, the
entanglement of $r_1r_2$ does not show ESB but remains to be zero.

\begin{figure}[tbp]
\begin{center}
\includegraphics[width = \columnwidth]{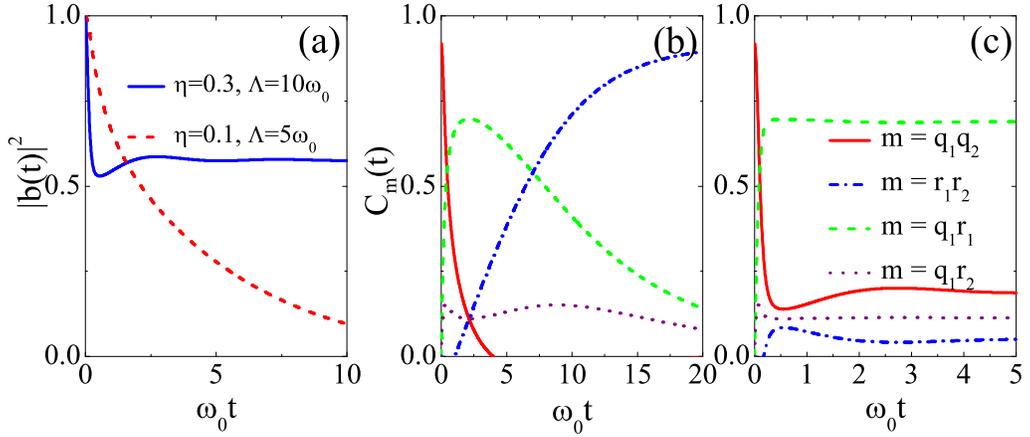}
\end{center}
\caption{(Color online) Entanglement evolution for the Ohmic
spectral density. The two sets of parameters $(\eta,\Lambda) = (0.1,
5\omega_0)$ and $(0.3, 10\omega_0)$ have been considered for
comparison. The corresponding entanglement distributions and their
evolution are given in (b) and (c), respectively. In both cases
$\alpha=0.55$.} \label{osd}
\end{figure}

The above discussion is general and is not dependent of the explicit
form of the reservoir. To confirm this, we consider the radiation
field in free space. The spectral density has the Ohmic form
$J(\omega)=\eta \omega \exp (-\omega/\Lambda)$, which can be
obtained from the free-space dispersion relation $\omega=ck$. One
can verify that the condition for the formation of bound states is:
$\omega_0-\eta \Lambda<0$ \cite{Tong09}. In Fig. \ref{osd}, we plot
the results in this situation. The previous results can be recovered
when the bound states are absent. On the contrary, when the bound
states are available, a stable entanglement is established among all
the bipartite partitions. Therefore, we argue that the stable
entanglement distribution resulted from the bound states is a
general phenomenon in open quantum system when the non-Markovian
effect is taken into account.

\section{Summary}
\label{conc} In summary, we have studied the entanglement
distribution among all the bipartite subsystems of two qubits
embedded into two independent reservoirs. It is found that the
entanglement can be stably distributed in all the bipartite
subsystems and they satisfy an identity about the entanglement. This
identity is shown to be independent of any detail of the reservoir
and its coupling with the qubit, which affect only the explicit time
evolution behavior and the final distribution. The result shows the
physical nature of the entanglement and has a significant
implication for the quantum information processing.

\chapter{Summary and outlook}\label{smotl}

In this thesis we studied the decoherence dynamics of open quantum
system. We found that the decoherence would be greatly suppressed if
the bound state is formed under the non-Markovian dynamics.

We model our system as two-level atoms in vacuum reservoirs. After
numerically solving the coupled equations, we studied the
non-Markovian effect on the decoherence dynamics of the quantum
system. Compared with the results obtained under the Markovian
approximation, we found that the environment has two effects on the
quantum system of interest: dissipation effect, which degrades the
quantum coherence, and backaction effect, which compensates the
quantum system of lost coherence. The competition of these two
effects results in the rich decoherence dynamic behaviors. Our
results show explicitly, in the weak coupling regime, the widely
used Markovian approximation is applicable, while, in the strong
coupling regime, the non-Markovian effect makes the dynamical
process oscillate for a certain time which is a manifestation of
information or energy flowing back and forth between quantum system
and the environment. We also find that the quantum coherence can be
preserved in the long time limit. Physically, it is attribute to the
formation of a bound state between the quantum system and its local
reservoir. We give explicitly a condition to judge the formation of
bound state for any kind of vacuum reservoir including the widely
studied PBG medium.

Due to the preservation of quantum coherence of the single system,
we have revealed that the entanglement for a composite two-qubit
system can also be preserved in the steady state. We give explicitly
the mechanism of entanglement preservation, i.e. the fulfillment of
non-Markovian effect and formation of bound states. The mechanism we
given can explain the results obtained in the previous works. When
the non-Markovian effect is neglected, the phenomenon of ESD of the
qubits is reproduced. The phenomenon of ESD and its revival can be
obtained when the non-Markovian effect is contained while the bound
state is not available. In particular, the entanglement preservation
when the atoms are placed in the PBG mediums reported in Ref.
\cite{Bellomo08} can be explained as the fulfillment of the above
two conditions. In a word, we have given a clear clue on how to
preserve entanglement in the steady state.

Considering the environment as a whole, we also investigated the
entanglement distribution among all the bipartite subsystems. We
found that the entanglement can be stably distributed among all
bipartite partitions of the whole system when the bound states are
formed. It is particularly interesting to find that the entanglement
in different bipartite partitions always satisfies an identity. This
identity is independent on the explicit dynamics process. Our
unified treatment includes the previous results in the literature as
special cases. When the bound state is absent and the Markovian
approximation is applicable, the result reported in Ref.
\cite{Lopez08} that the entanglement transfer from the qubits to the
reservoirs is recovered. Our work give a thorough understanding of
entanglement distribution among quantum systems and their
environments.

There are many open issues relevant to the subject of this thesis.
For example, the mechanism of formation of bound state is still
unclear, we think it is a kind of quantum phase transition when the
bound state is formed. How to reveal the relationship between the
bound state and the quantum phase transition is an open question.
What's more, in this thesis we have shown that the non-Markovian
effect can rescue the entanglement. Does the non-Markovian effect
also can rescue certain missions of quantum information processing,
for example, quantum teleportation and quantum dense coding, in
noisy quantum channels? This is still an open question.

{\small

\chapter*{Publication list}\parskip=-1mm \addcontentsline{toc}{chapter}{Publication list}

\begin{enumerate}

\item  {\bf Qing-Jun Tong},
Jun-Hong An, Hong-Gong Luo, and C. H. Oh, {\it Mechanism of
entanglement preservation}, Phys. Rev. A \textbf{81}, 052330 (2010).

\item Juan-Juan Chen, Jun-Hong An, {\bf Qing-Jun Tong}, Hong-Gang Luo, and C.
H. Oh, {\it Non-Markovian effect on the geometric phase of a
dissipative qubit}, Phys. Rev. A {\bf 81}, 022120 (2010).

\item {\bf Qing-Jun Tong}, Jun-Hong An, Hong-Gong Luo, and C. H. Oh, {\it Decoherence suppression of a dissipative qubit by non-Markovian
effect}, J. Phys. B: At. Mol. Opt. Phys. {\bf 43} 155501 (2010).

\item {\bf Qing-Jun Tong}, Jun-Hong An, Hong-Gang Luo, and C. H. Oh, {\it Entanglement distribution and its invariance}, arXiv:1005.1001, submitted.

\end{enumerate}}

\chapter*{Acknowledgments}\parskip=0.15cm
\addcontentsline{toc}{chapter}{Acknowledgments}\hskip1cm

As this thesis draws to a close, I give my great thanks to my
advisor, Professor Jun-Hong An for his patient instruction and
warm-hearted helps during the past years. Without his guidance and
help, it is impossible for me to go into the study of decoherence
dynamics of open quantum system and accomplish this thesis.

Furthermore, I express my great thanks to Professor Hong-Gang Luo
for his illuminating discussions and instructive helps. I also
acknowledge Dr. Zheng-Xiao Zhou, Dr. Lin Li, Juan-Juan Chen and many
other colleagues for their valuable helps and discussions during my
research works.

Finally, I can not forget the long-term support of my parents, my
brother and my lover who have shown patience and understanding to
the entire period of my work.

\end{document}